\documentclass[apj]{emulateapj}

\usepackage{natbib}
\usepackage{graphicx}
\usepackage{bm}
\usepackage{amsmath}
\usepackage{amsfonts}
\usepackage{amssymb}

\def\gtwid{\mathrel{\raise.3ex\hbox{$>$\kern-.75em\lower1ex\hbox{$\sim$}}}}
\def\ltwid{\mathrel{\raise.3ex\hbox{$<$\kern-.75em\lower1ex\hbox{$\sim$}}}}
\def\\{\hfil\break}

\newcommand{\R}{\textbf{R}\ }

\newcommand\T{\rule{0pt}{2.6ex}}       

\def\sun{\hbox{$\odot$}}

\def\lesssim{\mathrel{\hbox{\rlap{\hbox{\lower4pt\hbox{$\sim$}}}\hbox{$<$}}}}
\def\gtrsim{\mathrel{\hbox{\rlap{\hbox{\lower4pt\hbox{$\sim$}}}\hbox{$>$}}}}

\newcommand{\mamo}[1]{\mbox{$#1$}}
\newcommand{\unit}[1]{\ifmmode \:\mbox{\rm #1}\else \mbox{#1}\fi}
\newcommand{\mone}{\mamo{^{-1}}}

\newcommand{\kms}{\unit{km~s\mone}}

\newcommand{\kpc}{\unit{kpc}}

\newcommand{\LCDM}{$\Lambda$CDM}

\newcommand{\msun}{\mamo{M_{\sun}}}

\begin{document}

\title{Bayesian Mass Estimates of the Milky Way II: \\
The dark and light sides of parameter assumptions}
\author{Gwendolyn M. Eadie\altaffilmark{1}}
\author{William E. Harris\altaffilmark{1}}
\altaffiltext{1}{Dept.\ of Physics and Astronomy, McMaster University, Hamilton, ON L8S 4M1, Canada.}

\email{eadiegm@mcmaster.ca}
\shorttitle{}
\shortauthors{Eadie and Harris}

\begin{abstract}

We present mass and mass profile estimates for the Milky Way Galaxy using the Bayesian analysis developed by \cite{2015EHW} and using globular clusters (GCs) as tracers of the Galactic potential. The dark matter and GCs are assumed to follow different spatial distributions; we assume power-law model profiles and use the model distribution functions described in \cite{evans1997, deason2011, deason2012}. We explore the relationships between assumptions about model parameters and how these assumptions affect mass profile estimates. We also explore how using subsamples of the GC population beyond certain radii affect mass estimates. After exploring the posterior distributions of different parameter assumption scenarios, we conclude that a conservative estimate of the Galaxy's mass within 125\kpc~is $5.22\times10^{11}$\msun, with a 50\% probability region of $(4.79, 5.63) \times10^{11}$\msun. Extrapolating out to the virial radius, we obtain a virial mass for the Milky Way of $6.82\times10^{11}$\msun with 50\% credible region of $(6.06, 7.53) \times 10^{11}$\msun ($r_{vir}=185^{+7}_{-7}\kpc)$. If we consider only the GCs beyond 10\kpc, then the virial mass is $9.02~(5.69, 10.86) \times 10^{11}$\msun~ ($r_{vir}=198^{+19}_{-24}$\kpc). We also arrive at an estimate of the velocity anisotropy parameter $\beta$ of the GC population, which is $\beta=0.28$ with a 50\% credible region (0.21, 0.35). Interestingly, the mass estimates are sensitive to both the dark matter halo potential and visible matter tracer parameters, but are not very sensitive to the anisotropy parameter.

\end{abstract}


\maketitle

\label{firstpage}

\section{Introduction}\label{sec:intro}

The Milky Way's composition, structure, dynamical properties, and formation history are heavily influenced by two important properties: its total mass and mass profile. However,  inferring the mass profile of the Milky Way (MW) is a task fraught with uncertainty. Direct observations of dark matter, the most important component of the Galaxy's mass, still elude us. Therefore, astronomers must rely on kinematic information of tracer objects such as globular clusters (GCs), dwarf galaxies (DGs), stellar streams, and halo stars whose orbits are influenced by the Galaxy's gravitational potential.

Attempts to use these objects are made more difficult because many of the 3-dimensional velocity measurements are incomplete (i.e. proper motions are unknown). Although some tracers' velocity measurements are complete, popular mass estimators do not make use of both incomplete \emph{and} complete data at the same time. For example, projected mass estimators rely on line-of-sight velocities \citep[e.g. the mass estimators introduced by][]{bahcalltremaine1981, evanswilkinson2003, watkins2010}. On the other hand, mass estimators that use proper motions only use a subset of the data, because they require complete velocity vectors. So overall, we seem to have a dilemma; either we 1) throw away proper motion measurements, or 2) throw away some of the line-of-sight velocity measurements.

\cite{watkins2010} (hereafter W10) developed two different mass estimators: one that uses line-of-sight velocities only, and one that uses full 3-dimensional space motions. In an attempt to use all the data available, W10 employed each mass estimator separately, and combined the two estimates by a weighted average. This approach, however, requires a decision about how to weight each estimator, and also relies on Monte Carlo simulations to determine uncertainties. Furthermore, W10 find the mass estimators are quite sensitive to the velocity anisotropy.

Cosmological simulations have also shown that the spatial distributions of the dark matter and the tracers are probably quite different, and thus the density profiles of the tracers and the gravitational potential of the dark matter are not self-consistent. This inconsistency makes it difficult to model the phase-space distribution function of the tracer particles.

Because of the aforementioned issues, the MW's mass and mass profile estimates remain very uncertain, with values routinely varying between $10^{11}$ and $10^{12}$\msun \citep[see][for a graphic of mass estimates from studies using different methods]{wang2015}.

For quite some time now, in an effort to obtain better Galactic mass estimates, both maximum likelihood and Bayesian methods have been adopted, with the pioneering paper being \cite{little1987} \citep[see also][]{kulessa1992, kochanek1996, wilkinsonevans1999, mcmillan2011, kafle2012, 2015EHW, williamsevans2015, kupper2015}. 

\cite{2015EHW} (hereafter EHW) introduced a Bayesian analysis which estimates the cumulative mass profile of the Galaxy with both complete and incomplete kinematic data used simultaneously. The method uses the phase-space distribution function (DF) of the tracers, $f(\mathcal{E},L)$ as determined by the physical model \citep{binney2008}. Thorough testing of the method was performed with simulated data, and a preliminary analysis was done using real data (GCs and local DGs). The simulations showed that the Bayesian method is a powerful way to include complete and incomplete kinematic data simultaneously, and the preliminary analysis gave a total mass for the Galaxy that was in agreement with many other studies (although the range of values in the literature is wide). Furthermore, we found that estimating the mass in this way was relatively insensitive to the velocity anisotropy assumption.

In the discussion section of EHW, we listed ways to improve the analysis in future work. One major step-forward is to implement a model in which the spatial distribution of the dark matter halo is different from the spatial distribution of the satellites. Deriving a DF for such a model, in terms of the energies and momenta of tracers, can be quite difficult; the Eddington-equation method described in \cite{binney2008} (and also used below) requires the density profile of the tracers $\rho$ to be written as a function of the gravitational potential $\Phi$, which may not be possible if $\rho$ and $\Phi$ do not obey Poisson's equation. Still, there are realistic cases in which $\rho$ can be written as a function of $\Phi$, even when they are not self-consistent, and for which a non-negative DF can be found. 

\section{The Power Law Model}\label{sec:model}

We assume a galaxy model first proposed by \cite{evans1997}, and also used by \cite{deason2011, watkins2010}, in which the gravitational potential and the density profile of the satellites follow different power-law profiles. For such power-law profiles, the DF is analytic (see below), and thus provides a major advantage for assessing the effects of the important parameters of the model.

The first step in setting up the model is to derive the DF in terms of the parameters. \cite{evans1997} derived a family of DFs for their generalized power-law model in terms of the specific energy and angular momentum of galactic satellites. This DF was later adopted by \cite{deason2011, deason2012}, with a slightly different parametrization of the gravitational potential, and used in a maximum likelihood analysis to obtain an estimate for the mass profile of the Milky Way, given the kinematic information of blue horizontal branch (BHB).

Various pieces of the DF derivation are given in \cite{evans1997} and \cite{deason2011, deason2012}, but their notations differ substantially and the exact form of the normalization constant is unclear. In our experience, this can lead to confusion. Therefore, for completeness and clarity, we provide a short derivation of the DF using the parameterization and notation introduced by \cite{deason2011}.

The number density profile for the satellites (tracers) is given by
\begin{equation}\label{eq:rho}
	\rho_t \propto \frac{1}{r^{\alpha}}
\end{equation}
where $\alpha$ may be a free parameter (we drop the constant for Equation~\ref{eq:rho} as this factor is only related to the number of satellites). The gravitational potential of the dark matter halo (assumed to be spherical) is
\begin{equation}\label{eq:phi}
	\Phi = \frac{\Phi_o}{r^{\gamma}}
\end{equation}
where both $\gamma$ and $\Phi_o$ may be free parameters. The values of $\gamma$ and $\Phi_o$ will determine the cumulative mass profile of the dark matter halo through the equation:
\begin{equation}\label{eq:Mr}
	M(r) = \frac{\gamma\Phi_o}{G}\left(\frac{r}{\text{kpc}}\right)^{1-\gamma}
	\end{equation}
\citep{Deason2012ApJ}. For $\gamma \rightarrow 0$, $M(r)$ approaches an isothermal sphere. The opposite extreme, $\gamma \rightarrow 1$, corresponds to the Keplerian case of a central point mass. Mathematically, the values of $\Phi_o$ and $\gamma$ in equation~\ref{eq:Mr} may be any pair of real numbers. However, physically there are restrictions on their allowed values: $\Phi_o$ plays a large role in determining the mass and must be positive, and $0<\gamma<1$ if the cumulative mass profile is to be a constant or increasing function of radius. Although the parameters $\alpha$ and $\beta$ do not appear in Equation~\ref{eq:Mr}, their values will determine the shape of the posterior distribution, and may affect the estimates for $\Phi_o$ and $\gamma$.

We use the Eddington-equation method described in \cite{binney2008} to solve for the isotropic DF $f(\mathcal{E})$ in terms of the binding energy $\mathcal{E}= - v^2/2 + \Phi(r)$, where
\begin{equation}\label{eq:Eddington}
f(\mathcal{E})=\frac{1}{\sqrt{8}\pi^{2}}\int_{0}^{\mathcal{\mathcal{E}}}\frac{1}{\sqrt{\mathcal{E}-\Phi}}\left(\frac{d^{2}\rho_t}{d\Phi^{2}}\right)d\Phi+\frac{1}{\sqrt{\mathcal{E}}}\left(\frac{d\rho_t}{d\Phi}\right)_{\Phi=0}
\end{equation}
To solve Equation~\ref{eq:Eddington}, Equation~\ref{eq:rho} must be written as a function of Equation~\ref{eq:phi}. Solving Equation~\ref{eq:phi} for $r$, and substituting into Equation~\ref{eq:rho} gives
\begin{equation}
	\rho_t(\Phi) \propto \left(\frac{\Phi}{\Phi_o}\right)^{\alpha/\gamma}.
\end{equation}
The derivatives of the above equation are,
\begin{equation}\label{eq:first}
	\frac{d\rho_t}{d\Phi} \propto \frac{\alpha}{\gamma}\left(\frac{\Phi}{\Phi_o}\right)^{\frac{\alpha}{\gamma} - 1}\frac{1}{\Phi_o} 
\end{equation}
and
\begin{equation}\label{eq:second}
	\frac{d^2\rho_t}{d\Phi^2} \propto \frac{\alpha}{\gamma}\left(\frac{\alpha}{\gamma}-1\right)\left(\frac{\Phi}{\Phi_o}\right)^{\frac{\alpha}{\gamma} - 2}\frac{1}{\Phi_o^2}. 
\end{equation}
We assume and require that $\frac{\alpha}{\gamma}>1$ (i.e. the satellite profile is steeper than the dark matter profile). With this restriction in place, Equation~\ref{eq:first} goes to zero as $\Phi\rightarrow 0$, making the second term in Equation~\ref{eq:Eddington} vanish. The DF for the isotropic case is then,
\begin{equation}\label{eq:start}
	f(\mathcal{E})=\frac{1}{\sqrt{8}\pi^{2}}\frac{\alpha}{\gamma}\left(\frac{\alpha}{\gamma}-1\right)\frac{1}{\Phi_o^2} \int_{0}^{\mathcal{\mathcal{\mathcal{E}}}}\frac{1}{\sqrt{\mathcal{E}-\Phi}}\left(\frac{\Phi}{\Phi_o}\right)^{\frac{\alpha}{\gamma} - 2}d\Phi.
\end{equation}

The solution to this integral is analytic, albeit tedious. Use the substitution $u = \Phi/\mathcal{E}$ to solve the integral, and then apply the recursion relation $x\Gamma(x)=\Gamma(x+1)$, twice, to simplify the final expression to
\begin{equation}\label{eq:DFfinal}
	f(\mathcal{E}) = \frac{\mathcal{E}^{\frac{\alpha}{\gamma}-\frac{3}{2}}}{\sqrt{8\pi^3}\Phi_o^{\frac{\alpha}{\gamma}}}\frac{\Gamma\left( \frac{\alpha}{\gamma} + 1\right)}{\Gamma\left( \frac{\alpha}{\gamma} -\frac{1}{2}\right)}.
\end{equation}
Equation~\ref{eq:DFfinal} is the probability distribution of the specific energies of tracers in the potential $\Phi$, \emph{assuming an isotropic velocity dispersion}.

The velocity dispersion of Milky Way satellites is likely to be at least mildly anisotropic. \cite{1991cuddeford} showed that a way to incorporate velocity anisotropy is to multiply the DF by the specific angular momentum $L=rv_t$,
\begin{equation}\label{eq:DFL}
	f(\mathcal{E},L) \propto L^{-2\beta}f(\mathcal{E}).
\end{equation}
Here, $\beta$ is the velocity anisotropy parameter,
\begin{equation}\label{eq:beta}
	\beta = 1 - \frac{\sigma_{\theta}^2 + \sigma_{\phi}^2}{2\sigma_r^2}
\end{equation}
and $\sigma_{\theta}^2,~ \sigma_{\phi}^2, \text{and}~ \sigma_{r}^2$ are the velocity variances in spherical coordinates \citep{binney2008}. A system with completely radial orbits or completely tangential orbits will have $\beta=1$ or $\beta \rightarrow -\infty$ respectively, while an isotropic velocity dispersion will have $\beta=0$. 

\cite{evans1997} derived the normalization for Equation~\ref{eq:DFL}, and this factor depends on the parameters of interest. We reproduce their result here, the complete anisotropic DF, slightly re-organized and in Deason's notation:
\begin{equation}\label{eq:DFLfinal}
	f(\mathcal{E},L) = \frac{ L^{-2\beta}\mathcal{E}^{ \frac{\beta(\gamma-2)}{\gamma} + \frac{\alpha}{\gamma}-\frac{3}{2}} } {\sqrt{ 8\pi^{3} 2^{-2\beta}} \Phi_o^{-\frac{2\beta}{\gamma} + \frac{\alpha}{\gamma}}} \frac{
	\Gamma\left( \frac{\alpha}{\gamma} - \frac{2\beta}{\gamma}+ 1\right)}
	{\Gamma\left( \frac{\beta(\gamma-2)}{\gamma} + \frac{\alpha}{\gamma} -\frac{1}{2}\right)}.
\end{equation}
Note that as $\beta\rightarrow 0$, Equation~\ref{eq:DFLfinal} reduces to Equation~\ref{eq:DFfinal}.
  In summary, this model has four parameters:
\begin{itemize}
	\item[$\Phi_o$]{the scale factor for the gravitational potential}
	\item[$\gamma$]{the power-law slope of the gravitational potential}
	\item[$\alpha$]{the power-law slope of the satellite population}
	\item[$\beta$]{the velocity anisotropy parameter}
\end{itemize}
The parameters $\gamma,~ \alpha,~ \text{and} ~\beta$ are restricted by Equation~\ref{eq:DFLfinal} and the requirement that the DF be non-negative, 
\begin{equation}\label{eq:restrictions}
\alpha > \beta(2-\gamma) + \frac{\gamma}{2}
\end{equation}
\citep[][but be aware of notational differences]{evans1997}.

In practice, the dark matter halo profile is often assumed to follow an NFW-type or Sersic/Einasto-type function \citep[e.g.][]{merritt2006}. We experimented with such models but found that converting these to a DF becomes intractable analytically. There are numerical approximations to such models that are themselves quite complex \citep[e.g.][]{widrow2000}, but for the present purpose we stick to analytic models for simplicity. While a power-law potential for the dark matter is simplistic, it is also a common assumption in methodologies that use DFs \citep{deason2012, williamsevans2015}. Furthermore, a power-law potential (Equation~\ref{eq:phi}) has the benefit of approximating an NFW profile at large radii when $\gamma=0.5$ \citep{watkins2010, deason2011}, with recent analyses suggesting a transition radius around 10\kpc~\citep{huang2016arXiv, Harris2001book}.

In this study, we use the Deason power-law model shown in Equation~\ref{eq:DFLfinal}, and assume that all tracers are bound to the spherically symmetric, non-rotating system, i.e. $\mathcal{E}>0$. We explore the parameter space and mass profiles predicted by the model when it is confronted with real data.

\section{Kinematic Data: Globular Clusters}\label{sec:data}

In this study, and in contrast to EHW, we use only GCs to trace the Galactic potential. In principle, we could include kinematic data for both DGs and GCs to help extend the estimated $M(r)$ profile to larger distances. However, the model parameter $\alpha$ is meant to describe the power-law slope distribution of a \emph{single} population, and the GCs' and DGs' spatial distributions may be quite different. Thus, we only use GCs, but will return to this point in a later paper. Although metal-rich and metal-poor GCs may also have different distributions, we treat them as a single population: the metal-poor ones ([Fe/H]$<-1$) dominate the numbers, particularly at large Galactocentric radius.

Table~\ref{tab:deluxetable} lists all the kinematic data available for 157 Milky Way GCs, using the catalog of \cite{1996harris, 2010Harris} as a starting point. The Heliocentric line-of-sight velocities $v_{\text{los}}$ and the Galactocentric distances $r$ are from the Harris catalog, while the proper motions are taken from a variety of studies (see the ``$\mu$ Reference'' column in the table). The Galactocentric distances were calculated assuming the Sun's position with respect to the Galactic center as $(X_{\odot}, Y_{\odot}, Z_{\odot})=(8.0, 0, 0.02)$~kpc \citep[the height above the midplane is from][]{humphreys1995}. 

Almost half of the GCs in Table~\ref{tab:deluxetable} have measured proper motions, many of which are from the series of papers by Casseti-Dinescu, referenced collectively as ``Casseti'' in the table \citep{1999Dinescu, dinescu2004, dinescu2005, dinescu2010, dinescu2013}\footnote{Updated catalog: www.astro.yale.edu/dana/gc.html}. The GC M79 (NGC 1904) has two proper motion measurements we are aware of: the first was calculated by \cite{1999Dinescu} and is included in the Casetti online catalog, and the second was calculated by \cite{wangchenchen2005} using 29 years of photographic plates from the Shanghai Observatory. We use the result from \cite{wangchenchen2005} because it is more recent, and agrees well with \cite{1999Dinescu}.

Many Galactic GC proper motions are still unknown, although there are observational programs such as HSTPROMO (Sohn, S. et al 2016, in progress) which are beginning to remedy this issue.  For example, the proper motions of inner bulge GCs NGC 6522, NGC 6558, NGC 6540, NGC 6652, AL 3, ESO 456-SC38, Palomar 6, Terzan 2, Terzan 4, and Terzan 9 were recently measured by \cite{rossi2015}. A proper motion measurement for NGC 6681 was made for the first time by \cite{massari2013}, and an updated measurement for 47 Tuc (NGC 104) was recently completed by \cite{cioni2016}. 

Two GCs found in the bulge, NGC 6528 and NGC 6553, have proper motion measurements too \citep[][respectively]{feltzing2002, zoccali2001}, which can also be found in the Casseti online catalog. Pal 5's proper motion was measured by \cite{fritz2015} and \cite{kupper2015}, and their uncertainties overlap, so we use the average $\mathbf{\mu}$ of these two studies.

We exclude some GCs in Table~\ref{tab:deluxetable} from our analysis:
\begin{itemize}
  
  \item{A few objects simply do not have any velocity measurements, such as Ko 1, Ko 2, and AM 4.}
  
  \item{Some GCs are known to be associated with the Sagittarius DG, and their motions may be highly correlated with one another; these objects include Arp 2, NGC 6715, NGC 5634, Terzan 7, Terzan 8, and Whiting 1 \citep{law2010}.}
  
  \item{The following GCs are subject to very high reddening: NGC 6401, NGC 6544, Terzan 1, Pal 6, Djorg 1, and Terzan 6. However, because of the careful new measurements by \cite{rossi2015}, we include Terzan 1 and Pal 6 in the analysis.}
  
  \item{Another GC of issue is outer-halo object Pal 3; EHW showed that Pal 3's proper motion substantially affects the total mass estimate for the Milky Way. However, the  proper motion measurement for Pal 3 is highly uncertain \citep{majewski1993}. We are unaware of an updated proper motion measurement for Pal 3, and thus decide to treat the proper motion for this GC as unknown in the analysis.}
  
  \item{We exclude some GCs that do not have proper motions, and explain why below.}
 
\end{itemize}

To incorporate the incomplete data (GCs without proper motions) via the methodology of EHW, we must make the approximation that the Heliocentric line-of-sight velocity $|v_{\text{los}}|$ is approximately equal to the Galactocentric radial velocity $|v_r|$. This approximation is only valid when the GC is either 1) very far away, or 2) in line with the Sun and the Galactic center. To quantify this argument, we define $\xi$ as the angle subtended by the Sun, the GC, and the center of the Galaxy, and as in EHW require that GCs without a proper motion measurement have $|cos\xi|>0.95$ in order to be included. In these cases we can safely make the approximation that $|v_{\text{los}}|\approx |v_r|$.

\LongTables
\begin{deluxetable*}{lclrrrrc}[h]

\tablecaption{Kinematic Data of Galactic Globular Clusters\label{tab:deluxetable}}
\tablewidth{0pt}
\tablehead{
\colhead{Object} & {$r$} &\colhead{$\mu_{\alpha}\cos{\delta}$}     &\colhead{$\mu_{\delta}$}   & \colhead{$v_{\text{los}}$} & \colhead{$\cos{\xi}$} & \colhead{$\mathbf{\mu}$ Reference}  & \colhead{Included} \\
\colhead{ }   & \colhead{kpc}   &\colhead{(mas/year)}&\colhead{(mas/year)} & \colhead{\kms}  & \colhead{ }   & \colhead{}& \colhead{}     
}
\startdata
 NGC 104 & 7.4 & 7.26 $\pm$ 0.03 & -1.25 $\pm$ 0.03 & -18 $\pm$ 0.1 & 0.17 & Cioni & \checkmark \\ 
  NGC 288 & 12.0 & 4.675 $\pm$ 0.219 & -5.6 $\pm$ 0.35 & -45.4 $\pm$ 0.2 & 0.75 & Casseti & \checkmark \\ 
  NGC 362 & 9.4 & 4.873 $\pm$ 0.514 & -2.727 $\pm$ 0.824 & 223.5 $\pm$ 0.5 & 0.61 & Casseti & \checkmark \\ 
  Whiting 1 & 34.5 & --- & --- & -130.6 $\pm$ 1.8 & 0.98 & --- & --- \\ 
  NGC 1261 & 18.1 & --- & --- & 68.2 $\pm$ 4.6 & 0.90 & --- & --- \\ 
  Pal 1 & 17.2 & --- & --- & -82.8 $\pm$ 3.3 & 0.93 & --- & --- \\ 
  AM 1 & 124.6 & --- & --- & 116 $\pm$ 20 & 1.00 & --- & \checkmark \\ 
  Eridanus  & 95.0 & --- & --- & -23.6 $\pm$ 2.1 & 1.00 & --- & \checkmark \\ 
  Pal 2 & 35.0 & --- & --- & -133 $\pm$ 57 & 1.00 & --- & \checkmark \\ 
  NGC 1851 & 16.6 & 1.28 $\pm$ 0.68 & 2.39 $\pm$ 0.65 & 320.5 $\pm$ 0.6 & 0.89 & Casseti & \checkmark \\ 
  NGC 1904 & 18.8 & 2.34 $\pm$ 0.69 & -0.5 $\pm$ 0.75 & 205.8 $\pm$ 0.4 & 0.94 & Wang & \checkmark \\ 
  NGC 2298 & 15.8 & 4.05 $\pm$ 1 & -1.72 $\pm$ 0.98 & 148.9 $\pm$ 1.2 & 0.89 & Casseti & \checkmark \\ 
  NGC 2419 & 89.9 & --- & --- & -20.2 $\pm$ 0.5 & 1.00 & --- & \checkmark \\ 
  Ko 2 & 41.9 & --- & --- & --- & 1.00 & --- & --- \\ 
  Pyxis  & 41.4 & --- & --- & 34.3 $\pm$ 1.9 & 0.98 & --- & \checkmark \\ 
  NGC 2808 & 11.1 & 0.58 $\pm$ 0.45 & 2.06 $\pm$ 0.46 & 101.6 $\pm$ 0.7 & 0.71 & Casseti & \checkmark \\ 
  E 3 & 9.1 & --- & --- & --- & 0.57 & --- & --- \\ 
  Pal 3 & 95.7 & 0.33 $\pm$ 0.23 & 0.3 $\pm$ 0.31 & 83.4 $\pm$ 8.4 & 1.00 & Majewski \& Cudworth & \checkmark ($\mu$ not included) \\ 
  NGC 3201 & 8.8 & 5.28 $\pm$ 0.32 & -0.98 $\pm$ 0.33 & 494 $\pm$ 0.2 & 0.43 & Casseti & \checkmark \\ 
  Pal 4 & 111.2 & --- & --- & 74.5 $\pm$ 2.1 & 1.00 & --- & \checkmark \\ 
  Ko 1 & 49.3 & --- & --- & --- & 0.99 & --- & --- \\ 
  NGC 4147 & 21.4 & -1.54 $\pm$ 0.54 & -3.285 $\pm$ 0.516 & 183.2 $\pm$ 0.7 & 0.93 & Casseti & \checkmark \\ 
  NGC 4372 & 7.1 & -6.49 $\pm$ 0.33 & 3.71 $\pm$ 0.32 & 72.3 $\pm$ 1.2 & 0.24 & Casseti & \checkmark \\ 
  Rup 106 & 18.5 & --- & --- & -44 $\pm$ 3 & 0.93 & --- & --- \\ 
  NGC 4590 & 10.2 & -3.76 $\pm$ 0.66 & 1.79 $\pm$ 0.62 & -94.7 $\pm$ 0.2 & 0.70 & Casseti & \checkmark \\ 
  NGC 4833 & 7.0 & -8.11 $\pm$ 0.35 & -0.96 $\pm$ 0.34 & 200.2 $\pm$ 1.2 & 0.31 & Casseti & \checkmark \\ 
  NGC 5024 & 18.4 & 0.5 $\pm$ 1 & -0.1 $\pm$ 1 & -62.9 $\pm$ 0.3 & 0.90 & Casseti & \checkmark \\ 
  NGC 5053 & 17.8 & --- & --- & 44 $\pm$ 0.4 & 0.90 & --- & --- \\ 
  NGC 5139 & 6.4 & -5.08 $\pm$ 0.35 & -3.57 $\pm$ 0.34 & 232.1 $\pm$ 0.1 & 0.06 & Casseti & \checkmark \\ 
  NGC 5272 & 12.0 & -0.12 $\pm$ 0.607 & -2.667 $\pm$ 0.404 & -147.6 $\pm$ 0.2 & 0.75 & Casseti & \checkmark \\ 
  NGC 5286 & 8.9 & --- & --- & 57.4 $\pm$ 1.5 & 0.73 & --- & --- \\ 
  AM 4 & 27.8 & --- & --- & --- & 0.98 & --- & --- \\ 
  NGC 5466 & 16.3 & -3.9 $\pm$ 1 & 1 $\pm$ 1 & 110.7 $\pm$ 0.2 & 0.88 & Casseti & \checkmark \\ 
  NGC 5634 & 21.2 & --- & --- & -45.1 $\pm$ 6.6 & 0.96 & --- & --- \\ 
  NGC 5694 & 29.4 & --- & --- & -140.3 $\pm$ 0.8 & 0.98 & --- & \checkmark \\ 
  IC 4499 & 15.7 & --- & --- & 31.5 $\pm$ 0.2 & 0.91 & --- & --- \\ 
  NGC 5824 & 25.9 & --- & --- & -27.5 $\pm$ 1.5 & 0.98 & --- & \checkmark \\ 
  Pal 5 & 18.6 & -2.343 $\pm$ 0.356 & -2.3085 $\pm$ 0.331 & -58.7 $\pm$ 0.2 & 0.95 & Fritz; Kupper & \checkmark \\ 
  NGC 5897 & 7.4 & -4.93 $\pm$ 0.86 & -2.33 $\pm$ 0.84 & 101.5 $\pm$ 1 & 0.79 & Casseti & \checkmark \\ 
  NGC 5904 & 6.2 & 4.267 $\pm$ 0.597 & -11.3 $\pm$ 1.457 & 53.2 $\pm$ 0.4 & 0.33 & Casseti & \checkmark \\ 
  NGC 5927 & 4.6 & -5.72 $\pm$ 0.39 & -2.61 $\pm$ 0.4 & -107.5 $\pm$ 1 & 0.23 & Casseti & \checkmark \\ 
  NGC 5946 & 5.8 & --- & --- & 128.4 $\pm$ 1.8 & 0.67 & --- & --- \\ 
  BH 176 & 12.9 & --- & --- & --- & 0.94 & --- & --- \\ 
  NGC 5986 & 4.8 & -3.81 $\pm$ 0.45 & -2.99 $\pm$ 0.37 & 88.9 $\pm$ 3.7 & 0.67 & Casseti & \checkmark \\ 
  Lynga 7 & 4.3 & --- & --- & 8 $\pm$ 5 & 0.27 & --- & --- \\ 
  Pal 14 & 71.6 & --- & --- & 72.3 $\pm$ 0.2 & 1.00 & --- & \checkmark \\ 
  NGC 6093 & 3.8 & -3.31 $\pm$ 0.58 & -7.2 $\pm$ 0.67 & 8.2 $\pm$ 1.5 & 0.66 & Casseti & \checkmark \\ 
  NGC 6121 & 5.9 & -12.657 $\pm$ 0.285 & -19.387 $\pm$ 0.288 & 70.7 $\pm$ 0.2 & -0.94 & Casseti & \checkmark \\ 
  NGC 6101 & 11.2 & --- & --- & 361.4 $\pm$ 1.7 & 0.87 & --- & --- \\ 
  NGC 6144 & 2.7 & -3.06 $\pm$ 0.64 & -5.11 $\pm$ 0.72 & 193.8 $\pm$ 0.6 & 0.47 & Casseti & \checkmark \\ 
  NGC 6139 & 3.6 & --- & --- & 6.7 $\pm$ 6 & 0.70 & --- & --- \\ 
  Terzan 3 & 2.5 & --- & --- & -136.3 $\pm$ 0.7 & 0.23 & --- & --- \\ 
  NGC 6171 & 3.3 & -0.7 $\pm$ 0.9 & -3.1 $\pm$ 1 & -34.1 $\pm$ 0.3 & -0.29 & Casseti & \checkmark \\ 
  1636-283  & 2.1 & --- & --- & --- & 0.27 & --- & --- \\ 
  NGC 6205 & 8.4 & -0.103 $\pm$ 0.797 & 4.687 $\pm$ 0.813 & -244.2 $\pm$ 0.2 & 0.48 & Casseti & \checkmark \\ 
  NGC 6229 & 29.8 & --- & --- & -154.2 $\pm$ 7.6 & 0.97 & --- & \checkmark \\ 
  NGC 6218 & 4.5 & 1.15 $\pm$ 1.95 & -7.75 $\pm$ 1.672 & -41.4 $\pm$ 0.2 & -0.48 & Casseti & \checkmark \\ 
  FSR 1735 & 3.7 & --- & --- & --- & 0.63 & --- & --- \\ 
  NGC 6235 & 4.2 & --- & --- & 87.3 $\pm$ 3.4 & 0.89 & --- & --- \\ 
  NGC 6254 & 4.6 & -5.75 $\pm$ 0.778 & -4.75 $\pm$ 1.45 & 75.2 $\pm$ 0.7 & -0.58 & Casseti & \checkmark \\ 
  NGC 6256 & 3.0 & --- & --- & -101.4 $\pm$ 1.9 & 0.83 & --- & --- \\ 
  Pal 15 & 38.4 & --- & --- & 68.9 $\pm$ 1.1 & 0.99 & --- & \checkmark \\ 
  NGC 6266 & 1.7 & -3.5 $\pm$ 0.37 & -0.82 $\pm$ 0.37 & -70 $\pm$ 1.4 & -0.64 & Casseti & \checkmark \\ 
  NGC 6273 & 1.7 & -2.86 $\pm$ 0.49 & -0.45 $\pm$ 0.51 & 135 $\pm$ 4 & 0.55 & Casseti & \checkmark \\ 
  NGC 6284 & 7.5 & -3.66 $\pm$ 0.64 & -5.39 $\pm$ 0.83 & 27.6 $\pm$ 1.7 & 0.99 & Casseti & \checkmark \\ 
  NGC 6287 & 2.1 & -3.68 $\pm$ 0.88 & -3.54 $\pm$ 0.69 & -288.7 $\pm$ 3.5 & 0.73 & Casseti & \checkmark \\ 
  NGC 6293 & 1.9 & 0.26 $\pm$ 0.85 & -5.14 $\pm$ 0.71 & -146.2 $\pm$ 1.7 & 0.83 & Casseti & \checkmark \\ 
  NGC 6304 & 2.3 & -2.59 $\pm$ 0.29 & -1.56 $\pm$ 0.29 & -107.3 $\pm$ 3.6 & -0.88 & Casseti & \checkmark \\ 
  NGC 6316 & 2.6 & -2.42 $\pm$ 0.63 & -2.65 $\pm$ 0.56 & 71.5 $\pm$ 8.9 & 0.94 & Casseti & \checkmark \\ 
  NGC 6341 & 9.6 & -3.575 $\pm$ 0.893 & -0.6 $\pm$ 0.601 & -120 $\pm$ 0.1 & 0.61 & Casseti & \checkmark \\ 
  NGC 6325 & 1.1 & --- & --- & 29.8 $\pm$ 1.8 & -0.11 & --- & --- \\ 
  NGC 6333 & 1.7 & -0.57 $\pm$ 0.57 & -3.7 $\pm$ 0.5 & 229.1 $\pm$ 7 & 0.05 & Casseti & \checkmark \\ 
  NGC 6342 & 1.7 & -2.77 $\pm$ 0.71 & -5.84 $\pm$ 0.65 & 115.7 $\pm$ 1.4 & 0.39 & Casseti & \checkmark \\ 
  NGC 6356 & 7.5 & -3.14 $\pm$ 0.68 & -3.65 $\pm$ 0.53 & 27 $\pm$ 4.3 & 0.97 & Casseti & \checkmark \\ 
  NGC 6355 & 1.4 & --- & --- & -176.9 $\pm$ 7.1 & 0.88 & --- & --- \\ 
  NGC 6352 & 3.3 & --- & --- & -137 $\pm$ 1.1 & -0.59 & --- & --- \\ 
  IC 1257 & 17.9 & --- & --- & -140.2 $\pm$ 2.1 & 0.98 & --- & \checkmark \\ 
  Terzan 2 & 0.8 & -0.94 $\pm$ 0.3 & 0.15 $\pm$ 0.42 & 109 $\pm$ 15 & -0.59 & Rossi & \checkmark \\ 
  NGC 6366 & 5.0 & --- & --- & -122.2 $\pm$ 0.5 & -0.76 & --- & --- \\ 
  Terzan 4 & 1.0 & 3.5 $\pm$ 0.69 & 0.35 $\pm$ 0.58 & -50 $\pm$ 2.9 & -0.78 & Rossi & \checkmark \\ 
  HP 1 & 0.5 & --- & --- & 45.8 $\pm$ 0.7 & 0.43 & --- & --- \\ 
  NGC 6362 & 5.1 & -3.09 $\pm$ 0.46 & -3.83 $\pm$ 0.46 & -13.1 $\pm$ 0.6 & 0.26 & Casseti & \checkmark \\ 
  Liller 1 & 0.8 & --- & --- & 52 $\pm$ 15 & 0.30 & --- & --- \\ 
  NGC 6380 & 3.3 & --- & --- & -3.6 $\pm$ 2.5 & 0.91 & --- & --- \\ 
  Terzan 1 & 1.3 & 0.51 $\pm$ 0.31 & -0.93 $\pm$ 0.29 & 114 $\pm$ 14 & -1.00 & Rossi & \checkmark \\ 
  Ton 2 & 1.4 & --- & --- & -184.4 $\pm$ 2.2 & 0.23 & --- & --- \\ 
  NGC 6388 & 3.1 & -1.9 $\pm$ 0.45 & -3.83 $\pm$ 0.51 & 80.1 $\pm$ 0.8 & 0.71 & Casseti & \checkmark \\ 
  NGC 6402 & 4.0 & --- & --- & -66.1 $\pm$ 1.8 & 0.52 & --- & --- \\ 
  NGC 6401 & 2.7 & --- & --- & -65 $\pm$ 8.6 & 0.97 & --- & --- \\ 
  NGC 6397 & 6.0 & 3.69 $\pm$ 0.29 & -14.88 $\pm$ 0.26 & 18.8 $\pm$ 0.1 & -0.82 & Casseti & \checkmark \\ 
  Pal 6 & 2.2 & 2.95 $\pm$ 0.41 & 1.24 $\pm$ 0.19 & 181 $\pm$ 2.8 & -1.00 & Rossi & \checkmark \\ 
  NGC 6426 & 14.4 & --- & --- & -162 $\pm$ 23 & 0.96 & --- & \checkmark \\ 
  Djorg 1 & 5.7 & --- & --- & -362.4 $\pm$ 3.6 & 1.00 & --- & --- \\ 
  Terzan 5 & 1.2 & --- & --- & -93 $\pm$ 2 & -0.90 & --- & --- \\ 
  NGC 6440 & 1.3 & --- & --- & -76.6 $\pm$ 2.7 & 0.45 & --- & --- \\ 
  NGC 6441 & 3.9 & -2.86 $\pm$ 0.45 & -3.45 $\pm$ 0.76 & 16.5 $\pm$ 1 & 0.95 & Casseti & \checkmark \\ 
  Terzan 6 & 1.3 & --- & --- & 126 $\pm$ 15 & -0.91 & --- & --- \\ 
  NGC 6453 & 3.7 & --- & --- & -83.7 $\pm$ 8.3 & 0.98 & --- & \checkmark \\ 
  UKS 1 & 0.7 & --- & --- & 57 $\pm$ 6 & -0.24 & --- & --- \\ 
  NGC 6496 & 4.2 & --- & --- & -112.7 $\pm$ 5.7 & 0.86 & --- & --- \\ 
  Terzan 9 & 1.1 & 0 $\pm$ 0.38 & -3.07 $\pm$ 0.49 & 59 $\pm$ 10 & -0.79 & Rossi & \checkmark \\ 
  Djorg 2 & 1.8 & --- & --- & --- & -0.93 & --- & --- \\ 
  NGC 6517 & 4.2 & --- & --- & -39.6 $\pm$ 8 & 0.74 & --- & --- \\ 
  Terzan 10 & 2.3 & --- & --- & --- & -0.94 & --- & --- \\ 
  NGC 6522 & 0.6 & 3.35 $\pm$ 0.6 & -1.19 $\pm$ 0.34 & -21.1 $\pm$ 3.4 & -0.47 & Rossi & \checkmark \\ 
  NGC 6535 & 3.9 & --- & --- & -215.1 $\pm$ 0.5 & -0.05 & --- & --- \\ 
  NGC 6528 & 0.6 & -0.35 $\pm$ 0.23 & 0.27 $\pm$ 0.26 & 206.6 $\pm$ 1.4 & -0.13 & Feltzing \& Johnson & \checkmark \\ 
  NGC 6539 & 3.0 & --- & --- & 31 $\pm$ 1.7 & 0.12 & --- & --- \\ 
  NGC 6540 & 2.8 & 0.07 $\pm$ 0.4 & 1.9 $\pm$ 0.57 & -17.7 $\pm$ 1.4 & -0.95 & Rossi & \checkmark \\ 
  NGC 6544 & 5.1 & --- & --- & -27.3 $\pm$ 3.9 & -0.95 & --- & --- \\ 
  NGC 6541 & 2.1 & --- & --- & -158.7 $\pm$ 2.4 & -0.11 & --- & --- \\ 
  2MS-GC01  & 4.5 & --- & --- & --- & -0.95 & --- & --- \\ 
  ESO-SC06  & 14.0 & --- & --- & --- & 0.98 & --- & --- \\ 
  NGC 6553 & 2.2 & 2.5 $\pm$ 0.065 & 5.35 $\pm$ 0.076 & -3.2 $\pm$ 1.6 & -0.88 & Zoccali & \checkmark \\ 
  2MS-GC02  & 3.2 & --- & --- & -238 $\pm$ 36 & -0.95 & --- & --- \\ 
  NGC 6558 & 1.0 & -0.12 $\pm$ 0.55 & 0.47 $\pm$ 0.6 & -197.2 $\pm$ 1.5 & -0.56 & Rossi & \checkmark \\ 
  IC 1276 & 3.7 & --- & --- & 155.7 $\pm$ 1.3 & -0.53 & --- & --- \\ 
  Terzan 12 & 3.4 & --- & --- & 94.1 $\pm$ 1.5 & -0.90 & --- & --- \\ 
  NGC 6569 & 3.1 & --- & --- & -28.1 $\pm$ 5.6 & 0.95 & --- & \checkmark \\ 
  BH 261 & 1.7 & --- & --- & --- & -0.85 & --- & --- \\ 
  GLIMPSE02  & 3.0 & --- & --- & --- & -0.75 & --- & --- \\ 
  NGC 6584 & 7.0 & -0.22 $\pm$ 0.62 & -5.79 $\pm$ 0.67 & 222.9 $\pm$ 15 & 0.88 & Casseti & \checkmark \\ 
  NGC 6624 & 1.2 & --- & --- & 53.9 $\pm$ 0.6 & -0.01 & --- & --- \\ 
  NGC 6626 & 2.7 & 0.63 $\pm$ 0.67 & -8.46 $\pm$ 0.67 & 17 $\pm$ 1 & -0.89 & Casseti & \checkmark \\ 
  NGC 6638 & 2.2 & --- & --- & 18.1 $\pm$ 3.9 & 0.71 & --- & --- \\ 
  NGC 6637 & 1.7 & --- & --- & 39.9 $\pm$ 2.8 & 0.55 & --- & --- \\ 
  NGC 6642 & 1.7 & --- & --- & -57.1 $\pm$ 5.4 & 0.16 & --- & --- \\ 
  NGC 6652 & 2.7 & 4.75 $\pm$ 0.07 & -4.45 $\pm$ 0.1 & -111.7 $\pm$ 5.8 & 0.80 & Rossi & \checkmark \\ 
  NGC 6656 & 4.9 & 7.37 $\pm$ 0.5 & -3.95 $\pm$ 0.42 & -146.3 $\pm$ 0.2 & -0.95 & Casseti & \checkmark \\ 
  Pal 8 & 5.5 & --- & --- & -43 $\pm$ 15 & 0.92 & --- & --- \\ 
  NGC 6681 & 2.2 & 1.58 $\pm$ 0.18 & -4.57 $\pm$ 0.16 & 220.3 $\pm$ 0.9 & 0.55 & Massari & \checkmark \\ 
  GLIMPSE01  & 4.9 & --- & --- & --- & -0.54 & --- & --- \\ 
  NGC 6712 & 3.5 & 4.2 $\pm$ 0.4 & -2 $\pm$ 0.4 & -107.6 $\pm$ 0.5 & -0.09 & Casseti & \checkmark \\ 
  NGC 6715 & 18.9 & --- & --- & 141.3 $\pm$ 0.3 & 0.99 & --- & --- \\ 
  NGC 6717 & 2.4 & --- & --- & 22.8 $\pm$ 3.4 & -0.23 & --- & --- \\ 
  NGC 6723 & 2.6 & -0.17 $\pm$ 0.45 & -2.16 $\pm$ 0.5 & -94.5 $\pm$ 3.6 & 0.41 & Casseti & \checkmark \\ 
  NGC 6749 & 5.0 & --- & --- & -61.7 $\pm$ 2.9 & 0.30 & --- & --- \\ 
  NGC 6752 & 5.2 & -0.69 $\pm$ 0.42 & -2.85 $\pm$ 0.45 & -26.7 $\pm$ 0.2 & -0.50 & Casseti & \checkmark \\ 
  NGC 6760 & 4.8 & --- & --- & -27.5 $\pm$ 6.3 & 0.19 & --- & --- \\ 
  NGC 6779 & 9.2 & 0.3 $\pm$ 1 & 1.4 $\pm$ 1 & -135.7 $\pm$ 0.8 & 0.63 & Casseti & \checkmark \\ 
  Terzan 7 & 15.6 & --- & --- & 166 $\pm$ 4 & 0.98 & --- & --- \\ 
  Pal 10 & 6.4 & --- & --- & -31.7 $\pm$ 0.4 & 0.16 & --- & --- \\ 
  Arp 2 & 21.4 & --- & --- & 115 $\pm$ 10 & 0.99 & --- & --- \\ 
  NGC 6809 & 3.9 & -3.31 $\pm$ 0.945 & -9.695 $\pm$ 0.554 & 174.7 $\pm$ 0.3 & -0.47 & Casseti & \checkmark \\ 
  Terzan 8 & 19.4 & --- & --- & 130 $\pm$ 8 & 0.98 & --- & --- \\ 
  Pal 11 & 8.2 & --- & --- & -68 $\pm$ 10 & 0.83 & --- & --- \\ 
  NGC 6838 & 6.7 & -2.3 $\pm$ 0.8 & -5.1 $\pm$ 0.8 & -22.8 $\pm$ 0.2 & -0.06 & Casseti & \checkmark \\ 
  NGC 6864 & 14.7 & --- & --- & -189.3 $\pm$ 3.6 & 0.96 & --- & \checkmark \\ 
  NGC 6934 & 12.8 & 1.2 $\pm$ 1 & -5.1 $\pm$ 1 & -411.4 $\pm$ 1.6 & 0.86 & Casseti & \checkmark \\ 
  NGC 6981 & 12.9 & --- & --- & -345.1 $\pm$ 3.7 & 0.89 & --- & --- \\ 
  NGC 7006 & 38.5 & -0.96 $\pm$ 0.35 & -1.14 $\pm$ 0.4 & -384.1 $\pm$ 0.4 & 0.98 & Casseti & \checkmark \\ 
  NGC 7078 & 10.4 & -1.233 $\pm$ 0.617 & -7.567 $\pm$ 1.77 & -107 $\pm$ 0.2 & 0.70 & Casseti & \checkmark \\ 
  NGC 7089 & 10.4 & 5.9 $\pm$ 0.849 & -4.95 $\pm$ 0.849 & -5.3 $\pm$ 2 & 0.74 & Casseti & \checkmark \\ 
  NGC 7099 & 7.1 & 1.42 $\pm$ 0.69 & -7.71 $\pm$ 0.65 & -184.2 $\pm$ 0.2 & 0.45 & Casseti & \checkmark \\ 
  Pal 12 & 15.8 & -1.2 $\pm$ 0.3 & -4.21 $\pm$ 0.29 & 27.8 $\pm$ 1.5 & 0.91 & Casseti & \checkmark \\ 
  Pal 13 & 26.9 & 2.3 $\pm$ 0.26 & 0.27 $\pm$ 0.25 & 25.2 $\pm$ 0.3 & 0.95 & Casseti & \checkmark \\ 
  NGC 7492 & 25.3 & --- & --- & -177.5 $\pm$ 0.6 & 0.95 & --- & \checkmark  
\enddata
\end{deluxetable*}	

After culling the data for the above reasons, only 89 GCs remain in the sample.  We note that our main motivation for using the EHW Bayesian method was to incorporate all the available data, but we have just tossed aside over a third of the data, mostly due to our imposed geometric requirement that $|cos\xi|>0.95$. However, this is a temporary problem; in our next paper (Eadie et al 2016, in preparation), we will introduce a hierarchical version of the EHW method that negates the need for this geometric requirement. The GCs included in our current sample have a \checkmark~ in the ``Included"' column in  Table~\ref{tab:deluxetable}; 18 of the 89 GCs included in our sample have incomplete velocity measurements.

\section{Methods}
\subsection{Overview and Bayesian Inference}
We estimate the mass profile of the Milky Way's dark matter halo by assuming the power-law model described in Section~\ref{sec:model}, using the Bayesian method outlined in EHW, and confronting this model with the GC data described in Section~\ref{sec:data}. For numerical purposes, we use $G\equiv 1$ units. The cumulative mass profile in $10^{12}\msun$ units is then given by
\begin{equation}\label{eq:units}
	M(<r) = 2.325\times10^{-3}\gamma\Phi_o\left(\frac{r}{\text{kpc}}\right)^{1-\gamma}.
\end{equation}
where $\Phi_o$ has units $10^4\text{km}^2\text{s}^{-2}$ and $r$ is in \kpc.

The DFs in Eq.~\ref{eq:DFfinal} and \ref{eq:DFL} require Galactocentric velocities in a spherical coordinate system, rather than the Heliocentric proper motion and line-of-sight measurements presented in Table~\ref{tab:deluxetable}. In the Galactocentric spherical coordinate system, the binding energy $\mathcal{E}$ is given by
\begin{equation}\label{eq:energy}
	\mathcal{E} = -\frac{1}{2}(v^2_r + v^2_t) + \Phi(r)
\end{equation}
where $v_r$ and $v_t=\sqrt{v_{\phi}^2 + v^2_{\theta}}$ are the radial and tangential velocities respectively.

Heliocentric velocities $(v_{\text{los}}, \mu_\alpha\cos{\delta}, \mu_\delta)$ are transformed to Galactocentric velocities $(U,V,W)$ in a right-handed cylindrical coordinate system, following the method outlined in \cite{johnson1987}, but using J2000 epoch values for the North Galactic Pole. We assume the velocity of the Sun with respect to the local standard of rest is $(U_{\odot}, V_{\odot}, W_{\odot})=(11.1, 12.24, 7.25)$ \kms \citep{schonrich2010}, and take the local standard of rest velocity to be 220\kms. After transforming to $(U,V,W)$, the velocities are transformed to spherical coordinates $(v_r, v_{\phi}, v_{\theta})$.

When a tracer does not have a proper motion measurement, then the transformations described above cannot be computed. For these objects, $v_t$ in Equation~\ref{eq:energy} is still unknown. This is where using the Bayesian paradigm comes in handy: the unknown $v_t$'s can be treated as parameters in the model.

Bayes' theorem states that the posterior probability distribution $p(\bm{\theta}|\bm{y})$ is the probability of model parameters ($\bm{\theta}$), conditional on a set of data $\bm{y}$:
\begin{equation}
 p\left(\bm{\theta}|\bm{y}\right)=\frac{p\left(\bm{y}|\bm{\theta}\right)p\left(\bm{\theta}\right)}{p\left(\bm{y}\right)},
\label{eq:BayesTheorem}
\end{equation}
where $p\left(\bm{y}|\bm{\theta}\right)$ is the likelihood, and $p\left(\bm{\theta}\right)$ is the prior probability on $\bm{\theta}$ \citep{bayes1763}. The denominator is a normalization constant whose value is not of interest--- we may instead sample a distribution that is proportional to the posterior distribution,
\begin{equation}
 p\left(\bm{\theta}|\bm{y}\right) \propto p\left(\bm{y}|\bm{\theta}\right)p\left(\bm{\theta}\right),
\end{equation}
to obtain probabilities of model parameters given the data.

If there are $n$ tracers, each with data $(r, v_r, v_t)$, and assumed to be independent, then the posterior probability is proportional to the product 
\begin{align*}
 p\left(\bm{\theta}|\bm{y}\right) &\propto \prod_i^n p\left(y_i|\bm{\theta}\right)p\left(\bm{\theta}\right) \\
 &\propto \prod_i^n p\left((r_i, v_{r,i}, v_{t,i})|\bm{\theta}\right) p\left(\bm{\theta}\right).
\end{align*}
In the case that $v_t$ is unknown, it becomes a nuisance parameter in the model,
\begin{equation}\label{eq:post}
 p(\bm{\theta}|\bm{y}) \propto \prod_i^n p\left( (r_i, v_{r,i})|\bm{\theta}, v_{t,i}\right) p(v_{t,i}) p\left(\bm{\theta}\right).
\end{equation}
We define $p(v_{t,i})$, the prior probability on $v_{t,i}$, as a uniform distribution in $v_t^2$ (this accounts for spherical geometry). The nuisance parameters are sampled via a hybrid-Gibbs sampler, which is a mixture of a standard Metropolis algorithm \citep{metropolis1949,metropolis1953}  and a Gibbs sampler \citep{geman1984}. This method is an efficient way to treat the unknown tangential velocities as parameters; see \cite{eadieMSc2013} and EHW for more details. For a comprehensive description of Gibbs sampling, see \cite{gelman2003}.

\subsection{Markov Chains}

\begin{figure}
\centering             
	\includegraphics[trim=0cm 0.5cm 0cm 1.5cm, clip=true, totalheight=0.33\textheight]{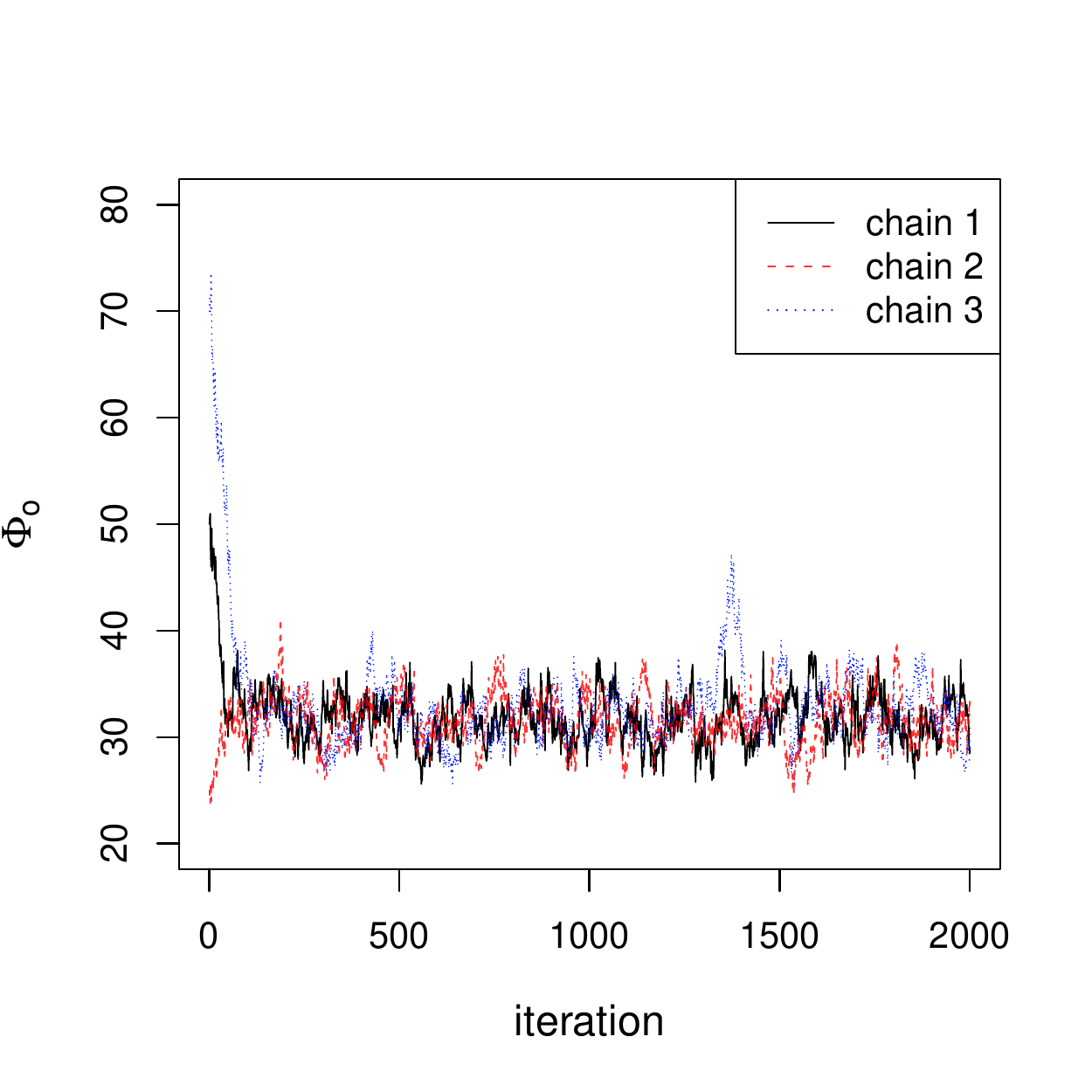}
	\caption{Example trace plot of three parallel Markov chains that are sampling the parameter space of $\Phi_o$.}
	\label{fig:traceplot}
\end{figure}

Samples of the posterior distribution are drawn via the hybrid-Gibbs sampler described in EHW.  We run three independent Markov chains in parallel: the  chains are initialized in different parts of parameter space and run until they reach a common stationary distribution (Figure~\ref{fig:traceplot}). The mutual convergence of the chains is assessed by inspecting the trace plots of the chains and by calculating the $\widehat{R}$ statistic \citep{gelman1992}.

Figure~\ref{fig:traceplot} is an example of a trace plot for three Markov chains that were initialized at different $\Phi_o$ values, but which have reached a common posterior distribution. Within the first few hundred iterations (the burn-in), these chains appear to have reached a common location in parameter space. The burn-in is discarded, and the Markov chains are run for at least $10^4$ more iterations, after which we confirm the effective sample size ($n_{\text{eff}}$) of the chains is at least 1000 (see EHW for a brief description of $n_\text{eff}$). After all requirements have been met (visual convergence, a $30-45\%$ acceptance rate, $\widehat{R}<1.1$, and $n_{\text{eff}}>1000$), we accept that the final Markov chain samples have  a distribution that is proportional to the posterior distribution (Eq. \ref{eq:post}). At this point, we calculate statistics, estimates, and probabilities of model parameters. 

\subsection{Priors}
Bayesian inference requires choosing prior probability distributions for model  parameters. We use flat, uniform prior probability distributions for three out of four model parameters, with the lower and upper bounds listed in Table~\ref{tab:priors}. When a parameter is held fixed, then the prior probability is a $\delta$-function centered on the chosen parameter value.

The prior on $\Phi_o$ is quite wide, representing little prior knowledge of the Galaxy's mass. In the cases of $\gamma$ and $\beta$, there are clear mathematical reasons for the prior bounds; $\gamma$ must be positive for the halo potential to decrease with distance (Equation~\ref{eq:phi}), and $\beta$ is the conventional anisotropy parameter, which has the limits described in Section~\ref{sec:model}.

Unfortunately, using $\beta\rightarrow-\infty$ as a lower-bound to a uniform distribution is ill-defined. On the other hand, there is strong evidence to expect $\beta \gtrsim 0$ for the Galactic stellar halo \citep{kafle2014ApJ}, and previous studies have shown that the velocity distribution of GCs in the Milky Way halo is mildly radial \citep{deason2011}. Furthermore, values of $\beta<-1$ are known to be unrealistic velocity anisotropies for distant halo stars \citep[e.g.][]{2016cunningham, deason2013halostars}. Taking all of this information into account, we set a conservative lower limit of $\beta_{\text{lower}}=-0.5$.

The parameter $\alpha$ determines the spherical density distribution of the GCs, and has been shown to follow a power-law profile with index $\sim3.5$ \citep{harris1976AJ,ejorgovski1994AJ}. Given our knowledge of the GC spatial distribution around the MW, it seems reasonable to define a slightly more informative prior distribution for this parameter. Although it may be tempting to look at all the GC $r$ values, fit a power-law profile, and then use the best-fit parameter value of $\alpha$ as a way to define a prior distribution, Bayesian inference can only use the data once; we cannot use the data to define a prior and then also use the same data to calculate the posterior distribution. This is especially true here, where the $r$ value of the GCs will help constrain not only $\alpha$, but also the potential through Equation~\ref{eq:energy}. So, how do we define a prior for $\alpha$?

We use the 68 MW GCs that are \emph{excluded} from our analysis (see Section~\ref{sec:data}) to determine the prior distribution for $\alpha$. The procedure to define the prior probability distribution $p(\alpha)$ is as follows.

First we hypothesize that the true density profile of the GCs is a power-law profile $\rho(r) \propto r^{-\alpha}$, implicitly assuming spherical symmetry.
We can re-write the power law as,
\begin{equation}\label{eq:powerlaw}
 \rho(r) \propto \frac{1}{4\pi r^2} \times \frac{1}{r^{\alpha-2}}.
\end{equation}
where the first term is one over the area of a sphere with radius $r$. Thus, in 1-dimension, the GCs follow a power-law with index $\alpha-2$. This kind of distribution can be described by a Pareto Distribution of the first kind.

The Pareto Distribution is a pdf defined as
\begin{equation}\label{eq:pareto}
 f(x|\eta, b) = \eta b^{\eta}~\frac{1}{x^{(\eta+1)}},~~0<b<x<\infty ,
\end{equation}
where $b$ is a threshold parameter--- the $x$ value beyond which the data follow a power-law  with index $\eta + 1$ \citep{2007Howlander, feigelson2012}. The term $\eta b^{\eta}$ is the normalization constant. Note the power-law slope in Equation~\ref{eq:pareto} is $\eta+1$, but in Equation~\ref{eq:powerlaw} is $\alpha-2$. Thus, for the GCs, $\alpha = \eta+3$.

Following \cite{2007Howlander}, the posterior pdf for $\eta$ given data vector $\textbf{x}$ of length $n$ is,
\begin{equation}\label{eq:posteta}
	p(\eta|b, \textbf{x}) = \frac{(nco+p)^{c+n}}{\Gamma(c+n)} \eta^{c+n-1} e^{-\eta (nc_o+p)}, \eta>0
\end{equation}
where $c$ and $p$ are parameters, and where
\begin{equation}
 nc_o = \sum^n_i \log\left( \frac{x_i}{b} \right).
\end{equation}
Equation~\ref{eq:posteta} is a Gamma distribution with shape and scale parameters ($n+c$) and ($nc_o+p$).

\begin{figure}
 \centering
	\includegraphics[scale=0.47]{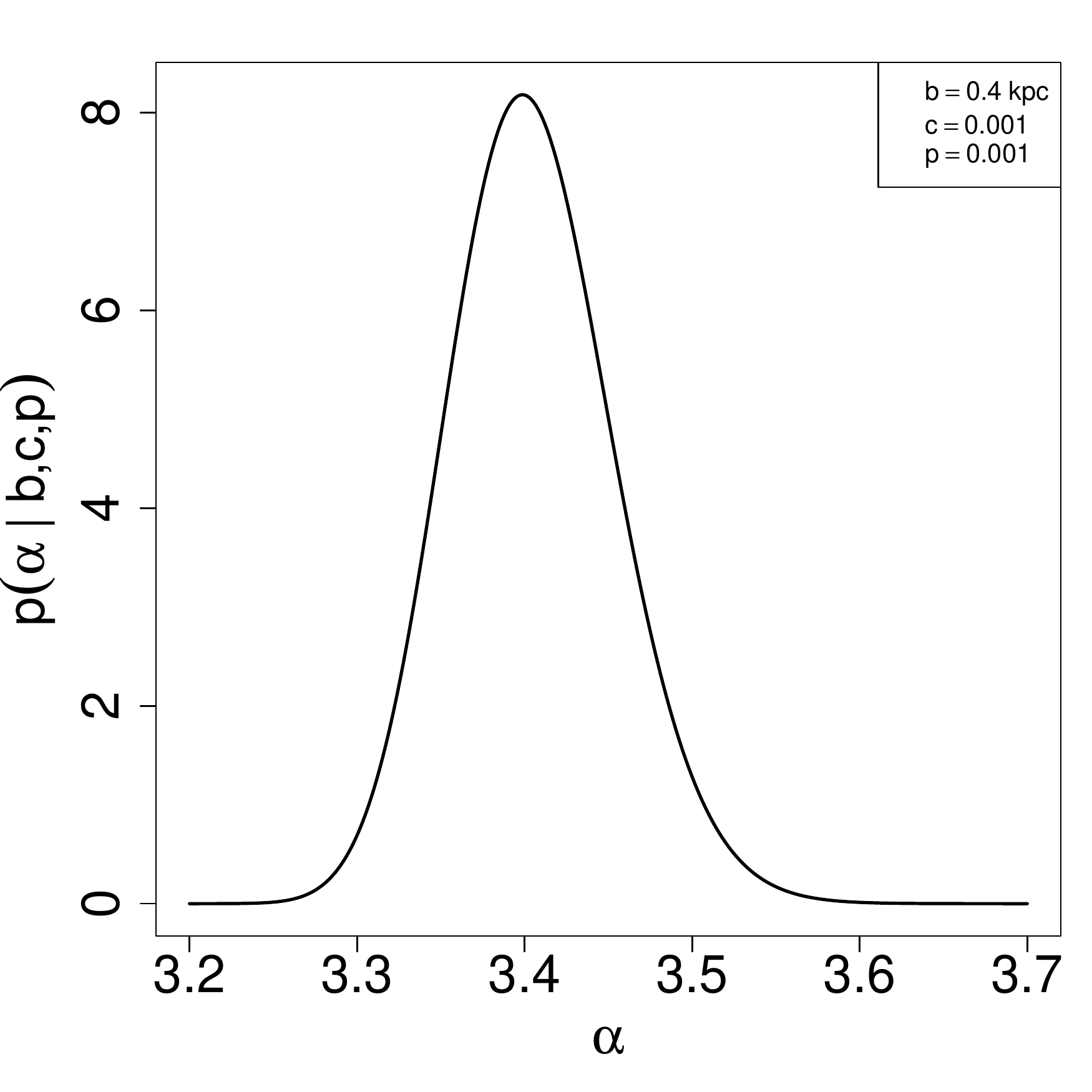}
	\caption{Prior probability distribution (a Gamma distribution) for $\alpha$, as determined by the extra GC data ($n=68$).}
	\label{fig:priora}
\end{figure}

We use Equation~\ref{eq:posteta} to calculate the probability distribution of $\eta$ given the extra data ($n=68$), and then reparameterize to obtain a prior probability distribution for $\alpha$ (Figure~\ref{fig:priora}). We let the extra GC data determine the shape and scale of the Gamma distribution as much as possible by defining $b=0.4$\kpc, $c=0.001$, and $p =0.001$. The most probable value in Figure~\ref{fig:priora} is $\alpha\approx3.4$. Interestingly, this value is in excellent agreement with the power-law best-fit obtained by \cite{wilkinsonevans1999}, who used a mixture of both globular clusters and dwarf galaxies beyond 20\kpc.
\begin{figure}[h]
\centering
	\includegraphics[scale=0.47]{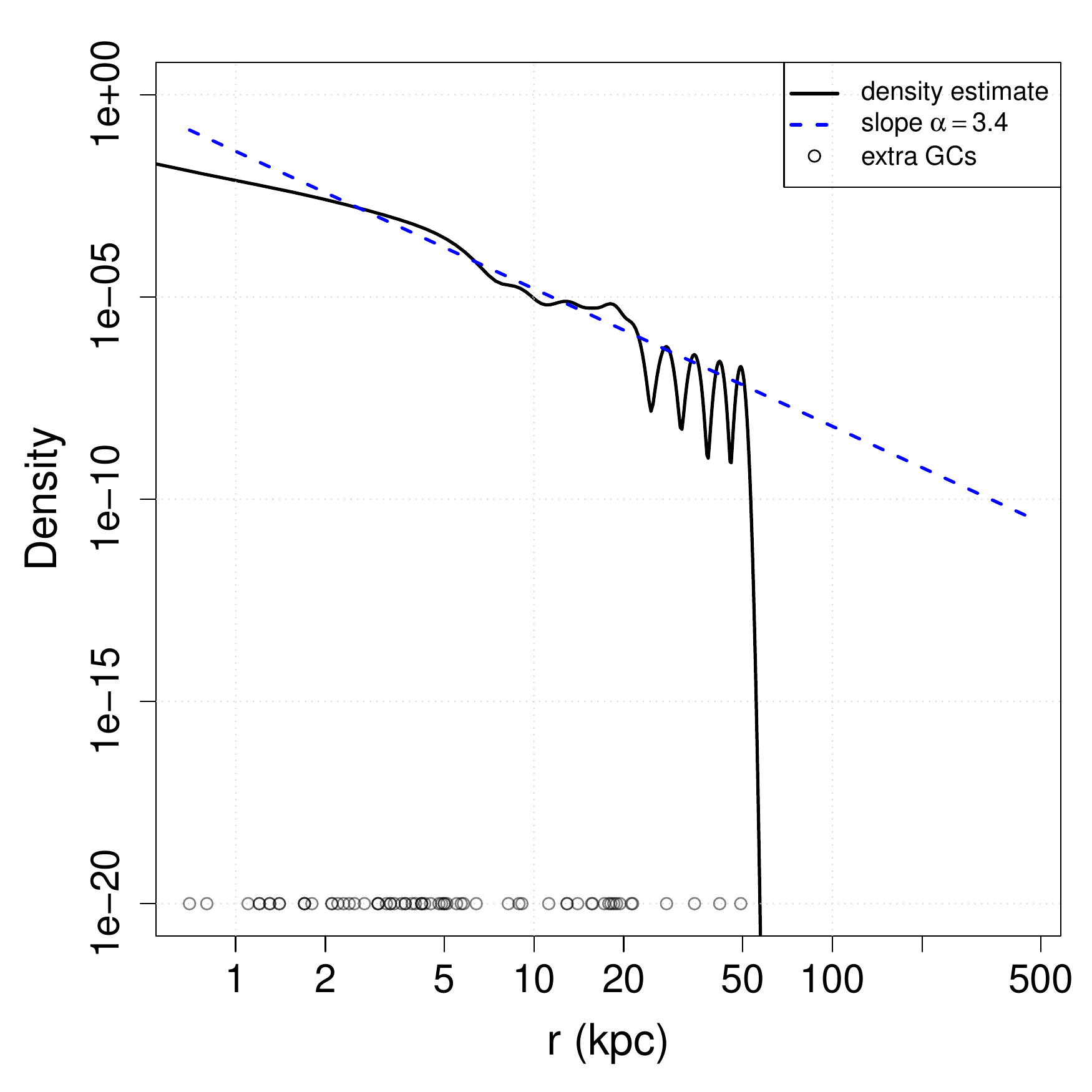}
	\caption{Density profile estimate of Milky Way Globular Clusters not used in this study. The dashed blue line has $\alpha=3.4$.}
	\label{fig:GCslope}
\end{figure}

As a check, we plot the smoothed density estimate of the extra GC data and a power-law profile with index 3.4 in Figure~\ref{fig:GCslope}. The smoothed density estimate is made using the \emph{density} function in the \R software environment, with a Gaussian kernel of bandwidth 1kpc. The power-law profile with slope 3.4 approximates the smoothed density quite well beyond 2\kpc. We will return to this point in Section~\ref{sec:789} below.

The spatial distribution of the 68 excluded GCs may be different from that of the 89 included GCs, due to selection effects (e.g. closer GCs may be more likely to have proper motion measurements). However, in a Bayesian analysis this is not a problem. By including the $r$ information of the excluded GCs via the prior, we are actually taking into account the spatial distribution of the \emph{entire} GC population. The estimate for $\alpha$ after using Bayes' theorem is a compilation of the information from the prior and the information from the data.

\begin{table}[t]
\centering
 \caption{Prior Probability Distributions for Parameters}
 \label{tab:priors}
 \begin{tabular}{ccc}
 \tableline
 \T
 Parameter & Prior & Prior Parameters \\
	\tableline
	\T
$\Phi_o$& Uniform & $\Phi_{o,\text{min}}=1$, $\Phi_{o,\text{max}}=200$ \\  
$\gamma$& Uniform  &$\gamma_{\text{min}}=0$, $\gamma_{\text{max}}=1$ \T \\
$\alpha$ & Gamma & $b=0.4\kpc$, $c=0.001$, $p=0.001$ \T \\
$\beta$ & Uniform &  $\beta_{\text{min}}=-0.5$, $\beta_{\text{max}}=1$ \T \\
 \tableline
 \end{tabular}
\end{table}

\subsection{Investigations}
The  $\Phi_o$ and $\gamma$ parameters directly determine the MW's mass profile (Equation~\ref{eq:Mr}), while parameters $\alpha$ and $\beta$ may indirectly affect it. Thus, a simple analysis would let only $\Phi_o$ and $\gamma$ be free parameters, while a more general analysis would let all four model parameters ($\Phi_o$, $\gamma$, $\alpha$, $\beta$) be free. To cover the range of possibilities, we instead perform an entire suite of investigations categorized first into Groups and then into Scenarios, as listed in Table~\ref{tab:Scenarios}.
\begin{table}[h]
\centering
 \caption{Investigations: fixed \emph{vs.} free parameters}
 \label{tab:Scenarios}
 \begin{tabular}{|l c|c|c|c|}
 \tableline
 \T  
\bf{Group} & \bf{Scenario} & \multicolumn{1}{|c|}{\bf{Potential}} & \multicolumn{1}{|c|}{\bf{Tracers}} & \multicolumn{1}{|c|}{\bf{Anisotropy}}\\
 \T
&  & $\gamma$ & $\alpha$ & $\beta$ \\
	\tableline
	\T
& I   &  &  & 0 \\
\bf{(1)} & II  & 0.5 & 3.5 & 0.5 \\
& III &     &     & \textbf{free} \\
\hline
& IV &       &     & 0  \\  
\bf{(2)} & V  &  \textbf{free} & 3.5 & 0.5 \\
& VI   &       &     & \textbf{free} \\
\hline
& VII &       &     & 0  \\  
\bf{(3)} & VIII & 0.5 & \textbf{free} & 0.5 \\
& IX  &       &     & \textbf{free}  \\ 
\hline
& X   &		  &     & 0  \\
\bf{(4)} & XI & \textbf{free} & \textbf{free} & 0.5 \\
& XII &       &     & \textbf{free}\\
 \tableline
 \end{tabular}
\end{table}

In every investigation, the parameter $\Phi_o$ is free. Group (1) holds $\gamma=0.5$ and $\alpha=3.5$ fixed, Group (2) holds $\alpha=3.5$ fixed, and Group (3) holds $\gamma=0.5$ fixed. Within each Group, we test Scenarios with different velocity anisotropies. For example, in Scenario IV, $\Phi_o$ and $\gamma$ are free, $\alpha$ is fixed at 3.5, and we assume an isotropic velocity dispersion ($\beta=0$). In Scenario VI, however, the anisotropy $\beta$ is a free parameter. Scenario XII is the most general analysis; $(\Phi_o, \gamma, \alpha, \beta)$ are all free.

\begin{table*}
	\centering
	 \caption{Summary of Parameter Estimates and 50\% marginal credible regions\label{tab:summary}}
	 \begin{tabular}{|l  c| cl | cl | cl | cl | cl | }
	 \tableline
	 \T
	    \textbf{Group } &~\textbf{Scenario}~ &
	    \multicolumn{2}{c|}{~$\Phi_o$ $(10^4 \text{km}^2\text{s}^{-2})$~} &
	    \multicolumn{2}{c|}{$\gamma$}&
	    \multicolumn{2}{c|}{$\alpha$}&
	    \multicolumn{2}{c|}{$\beta$}&
	    \multicolumn{2}{c|}{~~$M_{125}$ $(10^{11})$\msun~~} \\
		\tableline
		\T
&  I & 32.9 & (28.4-36.4) & 0.500 & --- & 3.500 & --- & 0.000 & --- & 4.28 & (3.69-4.73) \\ 
\textbf{(1)} &  II & 33.7 & (28.8-37.4) & 0.500 & --- & 3.500 & --- & 0.500 & --- & 4.38 & (3.75-4.86) \\ 
&  III & 33.4 & (28.6-37.0) & 0.500 & --- & 3.500 & --- & 0.353 & (0.291-0.422) & 4.34 & (3.72-4.81) \\
\hline 
&  IVb & 29.5 & (25.2-33.1) & 0.318 & (0.305-0.325) & 3.500 & --- & 0.000 & --- & 5.88 & (5.00-6.61) \\ 
\textbf{(2)} &  Vb & 30.4 & (25.7-34.3) & 0.329 & (0.309-0.342) & 3.500 & --- & 0.500 & --- & 5.93 & (4.97-6.72) \\ 
&  VIb & 29.5 & (27.2-31.5) & 0.333 & (0.310-0.348) & 3.500 & --- & 0.273 & (0.207-0.346) & 5.70 & (5.23-6.17) \\ 
\hline
&  VII & 31.4 & (27.4-34.5) & 0.500 & --- & 3.200 & (3.175-3.226) & 0.000 & --- & 4.09 & (3.56-4.48) \\ 
\textbf{(3)} &  VIII & 31.6 & (27.3-34.6) & 0.500 & --- & 3.200 & (3.174-3.225) & 0.500 & --- & 4.10 & (3.55-4.50) \\ 
&  IX & 31.5 & (27.4-34.6) & 0.500 & --- & 3.200 & (3.174-3.225) & 0.358 & (0.298-0.426) & 4.10 & (3.56-4.50) \\ 
\hline
&  X & 27.4 & (23.4-30.7) & 0.316 & (0.305-0.323) & 3.199 & (3.172-3.224) & 0.000 & --- & 5.47 & (4.66-6.13) \\ 
\textbf{(4)} &  XI & 27.4 & (23.1-30.9) & 0.327 & (0.309-0.339) & 3.198 & (3.171-3.224) & 0.500 & --- & 5.36 & (4.50-6.06) \\ 
&  XII & 26.9 & (24.9-28.8) & 0.332 & (0.309-0.346) & 3.199 & (3.175-3.221) & 0.276 & (0.21-0.349) & 5.22 & (4.79-5.63) \\ 
  
	\tableline
	
	 \end{tabular}
\end{table*}

\section{Results}

\begin{figure}[h]
\centering
	\includegraphics[scale=0.49]{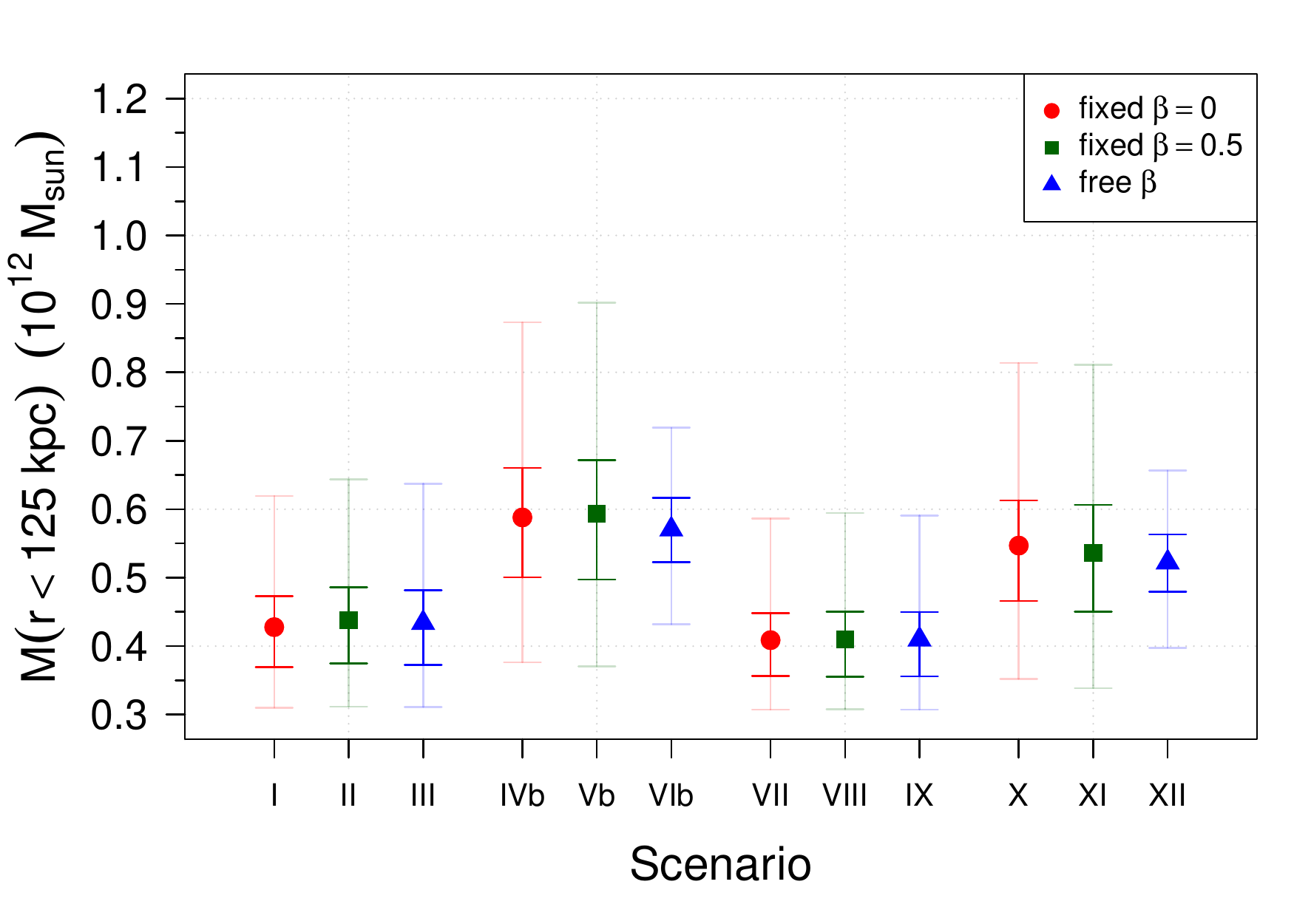}
	\caption{$M(r<125)$ estimates for all Scenarios, with 50\% and 95\% Bayesian credible intervals shown as bright and faint error bars respectively.}
	\label{fig:Mresults}
\end{figure}

Table~\ref{tab:summary} summarizes the results of our study, showing the estimates of each parameter and their 50\% marginal credible intervals in brackets. Figure~\ref{fig:Mresults} shows the mass estimates within 125\kpc, with error bars indicating the 50 and 95\% Bayesian credible intervals. The mass estimates are grouped as in Table~\ref{tab:Scenarios} and \ref{tab:summary} in order to highlight both the differences between groups, and differences between the anisotropy assumptions within a group. 

In the following four sections, the results are presented in more detail. Each section pertains to one of the Groups in Table~\ref{tab:Scenarios}. In this way, we intend to highlight the differences in mass estimates due to anisotropy assumptions vs. other parameter assumptions, and also describe why these differences occur.

\subsection{Group (1): Scenarios I, II, \& III}\label{sec:123}
Group (1) is the most rudimentary analysis, because it assumes that the dark matter profile is NFW-like in the outer halo, and that the power-law profile of the tracers is known with certainty.
\begin{figure*}
	\centering
	\includegraphics[trim=0cm 0cm 0cm 0cm, scale=0.85]{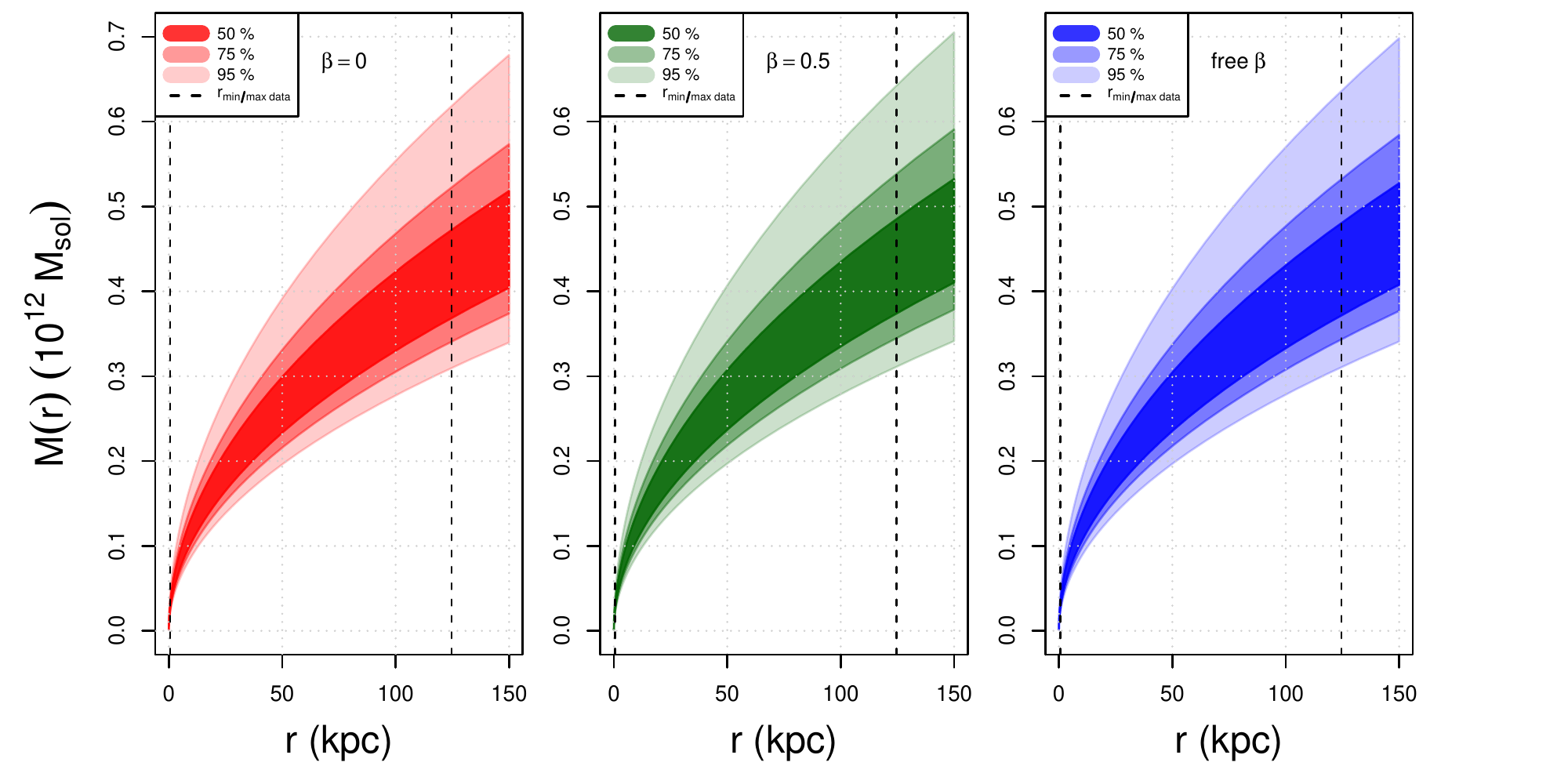}
	\caption{Credible regions for the cumulative mass profile in Scenarios I, II, and III. The percentages in the legend correspond to the three Bayesian credible regions shown, and the dotted lines show the extent of the GC data.}
	\label{fig:Mprofiles123}
\end{figure*}

The cumulative mass profile estimates for Scenarios I, II, \& III are presented in Figure~\ref{fig:Mprofiles123}, with the darkest regions representing the 50\% Bayesian credible regions. The velocity anisotropy assumption varies from left to right in the figure: the first assumes an isotropic velocity dispersion, the second a constant anisotropy of 0.5, and the third has $\beta$ as a free parameter.

Despite the variation in the anisotropy assumption between Scenarios I, II, and III, the mass profiles appear quite similar. The three estimates for the mass within 125\kpc~($M_{125}$) are 4.28, 4.38, and $4.34\times 10^{11}\msun$ and the 50\% probability credible regions are (3.69, 4.73), (3.75, 4.87), and (3.73, 4.82)$\times 10^{11}\msun$ respectively. The mass estimate in Scenario II ($\beta=0.5$) is only slightly higher than Scenarios in I ($\beta=0$) and III ($\beta$ free), but all estimates are in agreement within the 50\% credible regions.

For comparison, we also use W10's mass estimator (their Equation 24) with our complete data to compute a mass estimate, and then compare this result with our method. Using our data, $\alpha=3.5$, $\gamma=0.5$, and assuming an isotropic velocity assumption, W10's mass estimator returns  $1.79\times10^{11}$\msun~ for the mass within 38.5\kpc~ (the position of the outermost GC with a proper motion measurement). Under an anisotropic assumption of $\beta=0.5$, the W10 mass estimator gives $2.01\times10^{11}$\msun. In contrast, our cumulative mass profile and 50\% credible regions for Scenarios I and II are $M(r<38.5\kpc)=2.37~(2.05, 2.63)\times10^{11}\msun$, and $2.43~(2.08, 2.70)\times10^{11}\msun$. Therefore, the W10 mass estimator gives a slightly lower value using the complete data than our Bayesian method does using the complete and incomplete data together.
\begin{figure}
	\centering
	\includegraphics[scale=0.5]{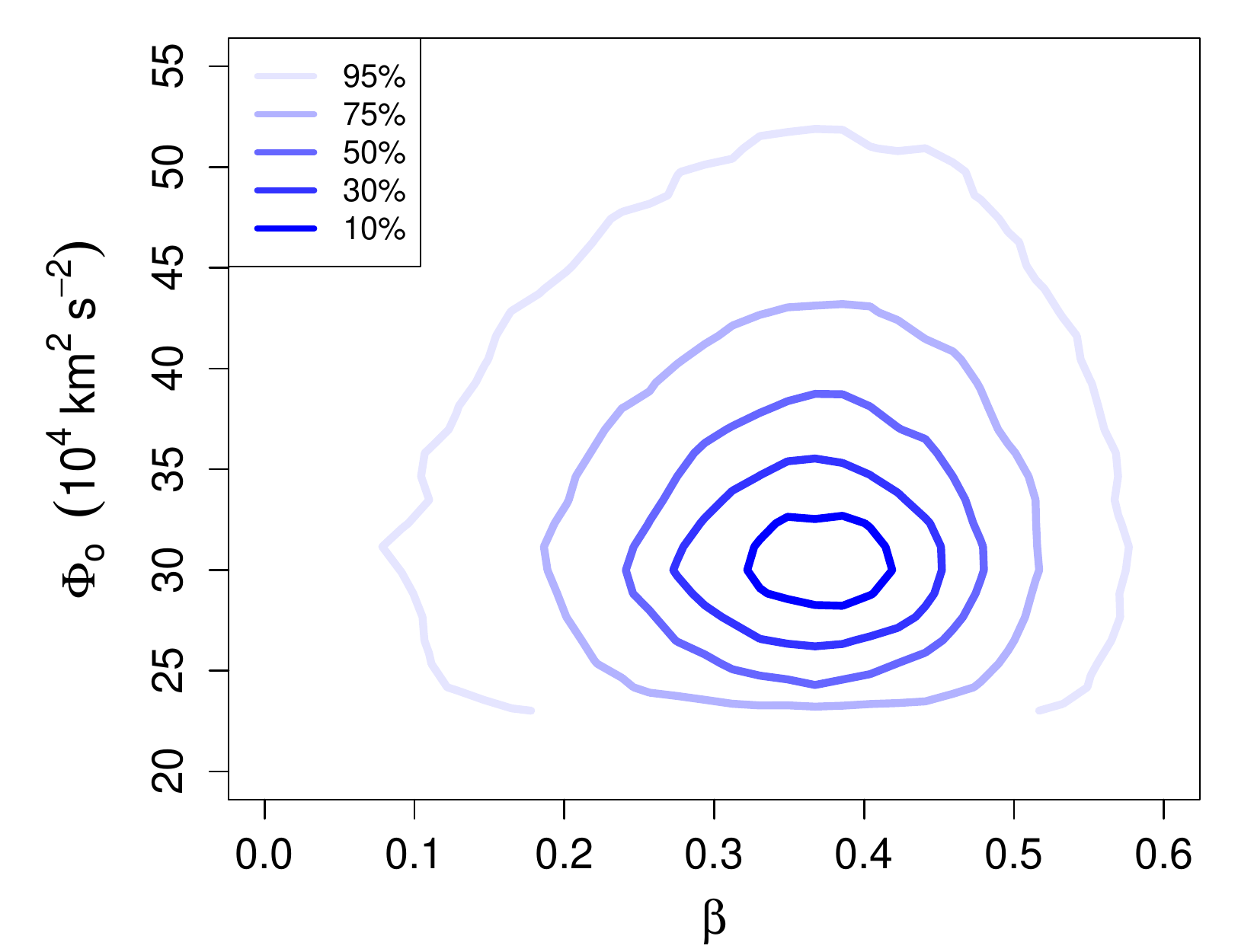}
	\caption{Posterior distribution for Scenario III. The contours show the 10, 30, 50, 75, and 95\% joint credible regions for $\Phi_o$ and $\beta$ when the parameters $alpha$ and $\gamma$ are fixed.}
	\label{fig:IIIpost}
\end{figure}

Figure~\ref{fig:IIIpost} shows Scenario III's 10, 30, 50, 75, and 95\% joint credible regions for $\Phi_o$ and $\beta$ \footnote{Contours are drawn with the \emph{emdbook} package in \textbf{R} \cite[see][]{emdbookBOOK, emdbook1}}. There is no evidence for strong correlation between the two parameters, and the estimate of $\beta$ suggests a slightly anisotropic velocity distribution. The 50\% marginal credible interval for $\beta$ is between 0.29 and 0.42, and the mean estimate for $\beta$ is 0.35. Note that the 50\% \emph{joint} credible interval for $\beta$ and $\Phi_o$ has a slightly wider range than the \emph{marginal} credible interval for $\beta$ alone.

\subsection{Group (2): Scenarios IV, V, \& VI}\label{sec:456}
Scenarios IV - VI investigate the case in which $\gamma$ is a free parameter, constrained to the lower and upper bounds given in Table~\ref{tab:priors}. In principle, this Group is the one we should pay most attention to, because $\gamma$ is the least constrained by observations. 

\begin{figure}
	\includegraphics[scale=0.5]{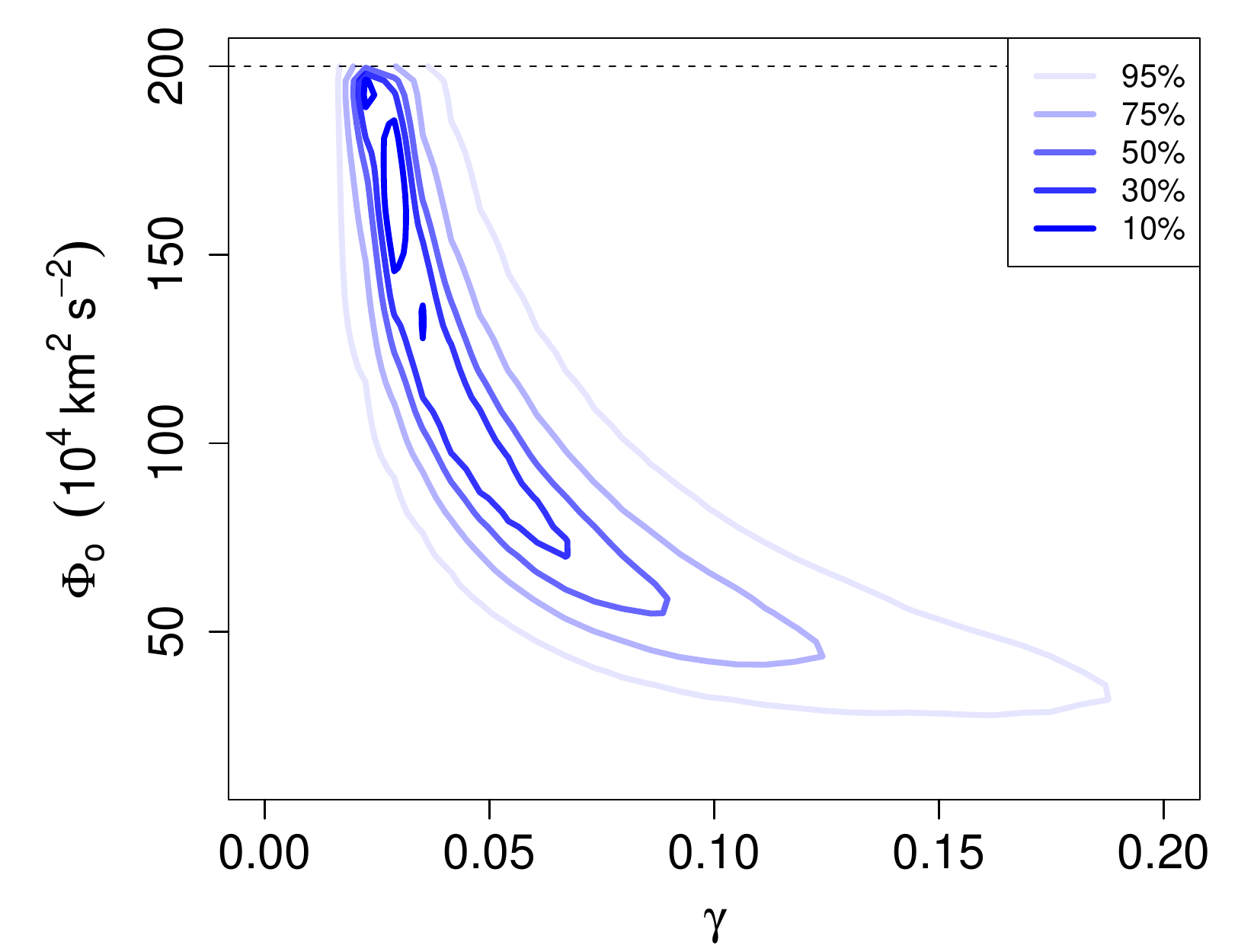}
	\caption{\textbf{Scenario IV posterior distribution for $\Phi_o$ and $\gamma$}. The contours show the Bayesian joint credible region. The dotted horizontal line is the upper bound of the uniform prior on $\Phi_o$. }
	\label{fig:2dpost}
\end{figure}

Figure~\ref{fig:2dpost} shows the joint credible regions for $\Phi_o$ and $\gamma$ in Scenario IV, after assessing for convergence. Even though $\gamma$ was allowed to vary between 0 and 1 (a flat, uniform prior), very few samples are drawn from the region $\gamma>0.3$. The parameters $\Phi_o$ and $\gamma$ have a strong, highly non-linear correlation at low $\gamma$ values, which results in larger estimates for the mass. The shape of the distribution is also reminiscent of that seen in \cite{deason2012ApJellipticals}. The mean value of $\Phi_o$ from the posterior distribution is $102.7$, with a 50\% credible region of (64.2, 138.1)$\times10^4\text{km}^{2}\text{s}^{-2}$, and the mean estimate for $\gamma$ is 0.06 (the median is 0.05), with 50\% marginal distribution samples between 0.03 and 0.08. The latter estimate is quite surprising, considering that \cite{deason2012} found $\gamma\sim0.35$ using the same model and BHB stars as tracers.

When the mean $\Phi_o$ and $\gamma$ from the posterior distribution in Figure~\ref{fig:2dpost} are  naively used as the best estimates for these parameters, then the cumulative mass profile increases almost linearly with distance from the Galactic center, as it should for $\gamma \rightarrow 0$ (Equation~\ref{eq:Mr}). The mean mass estimate within 125\kpc~ is also significantly larger than in Group (1), at a value of $1.09\times10^{12}\msun$, with a 50\% credible region of (0.92, 1.24)$\times10^{12}\msun$. 

A numerical issue is that the posterior distribution of $\Phi_o$ and $\gamma$ in Scenario IV may be multi-modal, and that the Markov chains may be stuck in a local maximum. Figure~\ref{fig:2dpost} shows more than one peak in the posterior distribution which may be real modes as opposed to numerical artifacts, ($10^6$ pairs of $\Phi_o$ and $\gamma$ were drawn in Scenario IV). One could use a more complicated sampling method to try to explore other parts of the parameters space \citep[e.g. the affine invariant sampler introduced by][]{goodman2010}, but we view this as unnecessary since we have good reasons to put a narrower prior distribution on $\gamma$.

An isothermal profile for the dark matter halo ($\gamma\rightarrow0$ in Equation~\ref{eq:Mr}) has been ruled out in the case of constant anisotropy \citep{battaglia2005}. At the other extreme,  $\gamma\rightarrow1$, $M(r)$ goes to a point mass, which is unrealistic for a dark matter halo. As stated previously, $\gamma=0.5$ is a good approximation to an outer NFW-type dark matter halo, for galaxies like the Milky Way \citep{watkins2010,deason2011}. This is why we choose $\gamma=0.5$ in Scenarios where $\gamma$ is fixed. In the scenarios where $\gamma$ is a free parameter, we relax the NFW approximation slightly and apply a uniform prior with lower and upper bounds of 0.3 and 0.7 respectively, calling these Scenarios IVb, Vb, and VIb. The range $\gamma=0.3$ to 0.7 by itself covers a large range in dark matter halo central concentrations, while staying within arguably realistic bounds.

The above adjustment to the prior $p(\gamma)$ allows for slightly non-NFW type potentials while notably excluding some parameter ranges. Because of the relationship seen in Figure~\ref{fig:2dpost}, the new prior on $\gamma$ \emph{will} change our mass estimates. However, choosing this slightly informative prior is important in the Bayesian paradigm--- we apply prior distributions based on our assumptions and current knowledge about the situation, including external information not contained in the GC data themselves. At the same time, we  present our results with the reminder that they are influenced by our assumptions, and thus are open to interpretation and criticism.

\begin{figure}[t]
	\centering
	\includegraphics[scale=0.45]{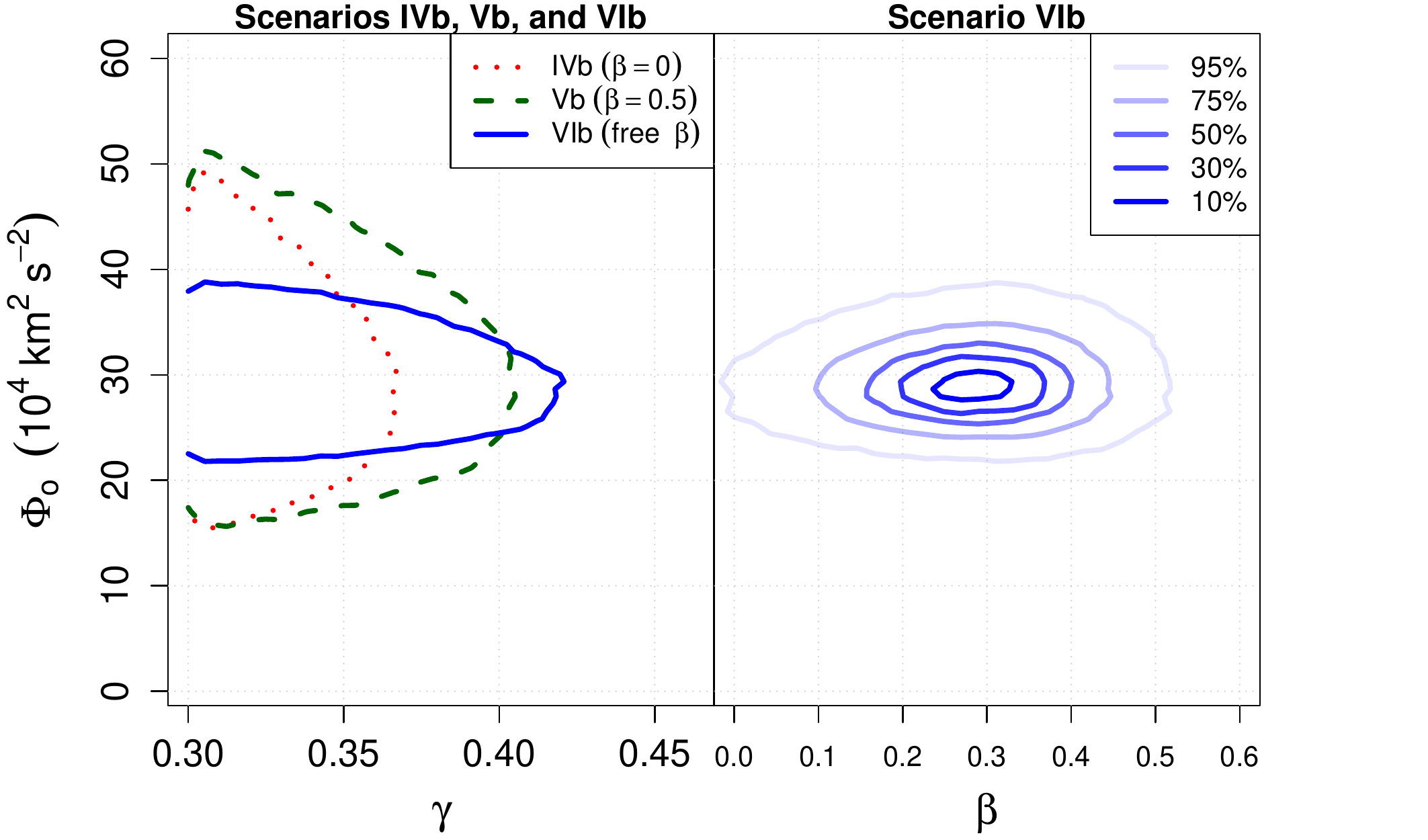}
	\caption{\textbf{Left:} The 95\% joint credible regions for Group (2). The regions for the isotropic ($\beta=0$), constant anisotropic ($\beta=0.5$), and constant anisotropic (free parameter) are shown in dotted (red), dashed (green), and solid (blue) lines respectively. \textbf{Right:} The posterior distribution and joint credible regions for $\Phi_o$ and $\gamma$ in Scenario VIb.}
	\label{fig:456b-post}
\end{figure}

With the new prior on $\gamma$, the mean estimates for Scenario IVb are  of $\Phi_o=29.5$ (25.2, 33.1)$\times10^4\text{km}^{2}\text{s}^{-2}$ and $\gamma=0.318$ (0.305, 0.325). These estimates are in better agreement with \cite{deason2012}, where the same model was applied to BHB stars using a maximum likelihood method rather than a Bayesian analysis. The mean estimates of $\Phi_o$ and $\gamma$ in Scenarios Vb and VIb are very similar, but the shapes of the 95\% joint posterior probability contours are rather different, shown in Figure~\ref{fig:456b-post}. Notice that when $\beta$ is a free parameter, the range of $\Phi_o$ is considerably smaller. Although it may initially seem strange that an extra free parameter would cause the posterior distribution to narrow, one explanation is that the posterior distribution of $\Phi_o$ and $\beta$ is narrow as well (the right panel of Figure~\ref{fig:456b-post}).

\begin{figure*}
	\centering
	\includegraphics[trim=0cm 0cm 0cm 0cm, scale=0.75]{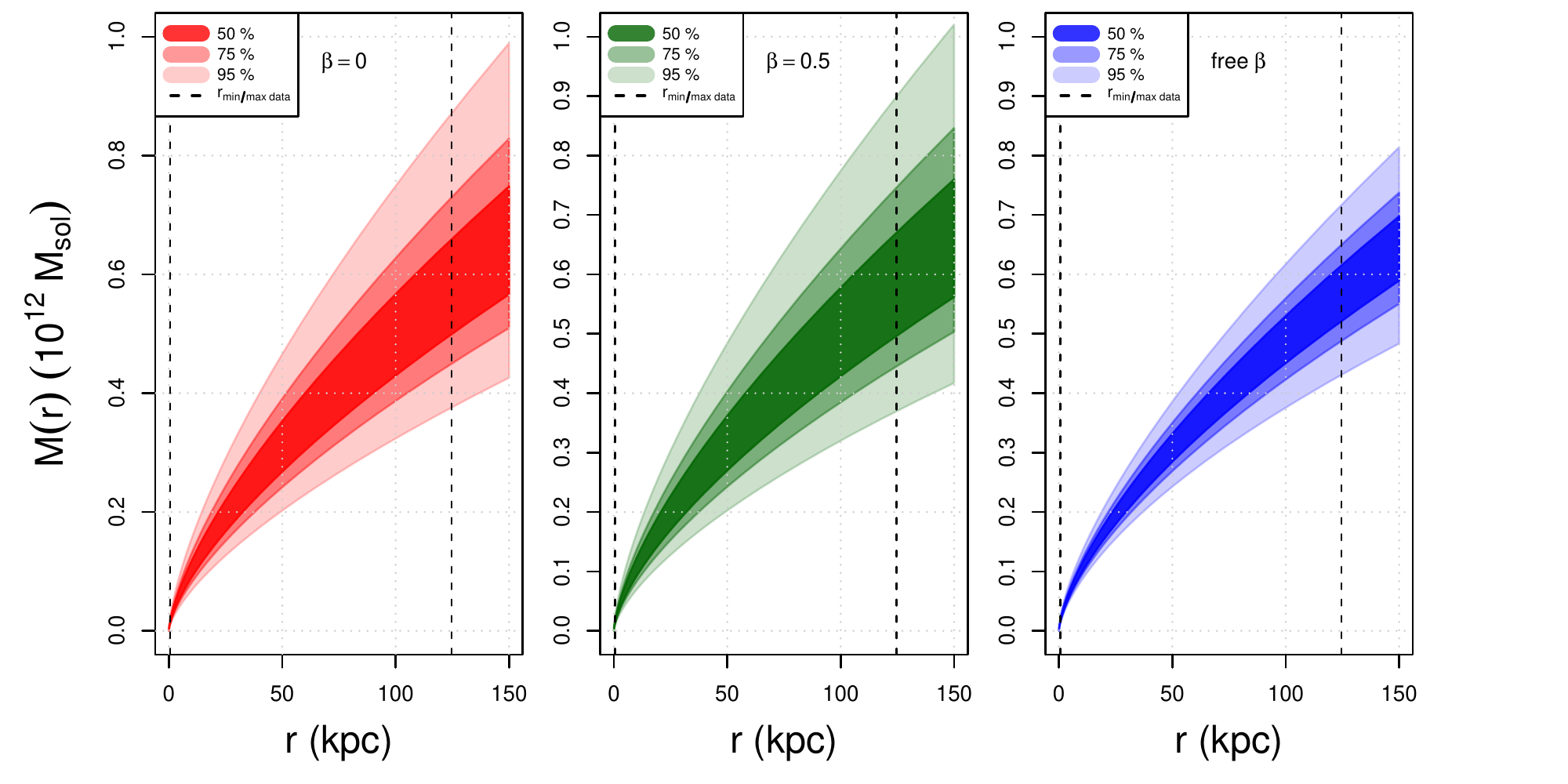}
	\caption{\textbf{Credible regions for the cumulative mass profile in Scenarios IVb, Vb, and VIb}. The percentages in the legend correspond to the three Bayesian credible regions shown, and the dotted lines show the extent of the GC data.}
	\label{fig:Mprofiles456}
\end{figure*}

The $M_{125}$ estimates for Scenarios IVb, Vb, and VIb are considerably higher than those in Scenarios I, II, and III, but the Bayesian credible intervals are also substantially larger (Figure~\ref{fig:Mresults} and Table~\ref{tab:summary}). Likewise, the cumulative mass profile credible regions are also substantially wider (Figure~\ref{fig:Mprofiles456}). In the isotropic and anisotropic cases (Scenario IVb and Vb), the 50\% credible regions for $M_{125}$ are $5.00-6.61\times10^{11}\msun$ and $4.97-6.72\times10^{11}\msun$ respectively. The $M_{125}$ estimate is most constrained in Scenario VIb (free $\beta$) with a 50\% credible interval of $5.23-6.17\times10^{11}\msun$. This is attributed to the narrowed marginal  distribution of $\Phi_o$ when $\beta$ is a free parameter (Figure~\ref{fig:456b-post}).

The differences in mass estimates between Group (1) and Group (2) are significant in terms of the 50\% credible intervals. The 95\% credible intervals for Scenarios IVb, Vb, and VbI contain the mass estimates obtained in Group (1), although VIb is a close call. The differences between Groups (1) and (2) must arise because $\gamma$ is a free parameter. The estimate of $\gamma$ is near the lower bound of the prior distribution $p(\gamma)$, which might indicate that the prior is too strong of an assumption. On the other hand, the uncertainty in the mass is very large when $\gamma$ is free, suggesting that there is insufficient information in the GC data alone to pin down the shape of the dark matter halo.

\subsection{Group (3): Scenarios VII, VIII, \& IX}\label{sec:789}

\begin{figure}
	\centering
	\includegraphics[scale=0.37]{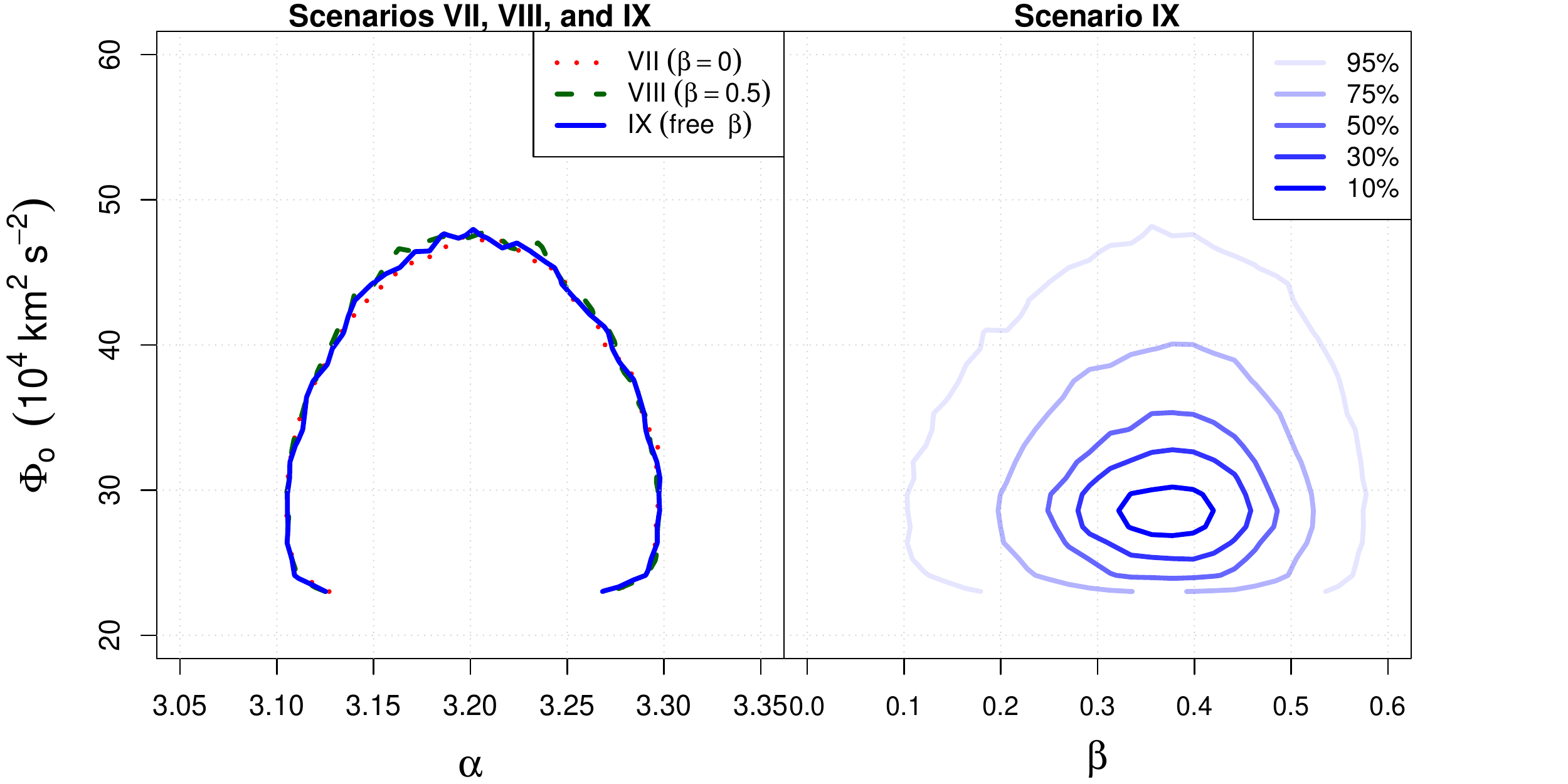}
	\caption{\textbf{Left:} The 95\% joint credible region contours for Scenarios VII, VIII, and IX. The regions for the isotropic ($\beta=0$), constant anisotropic ($\beta=0.5$), and constant anisotropic (free parameter) are shown in dotted (red), dashed (green), and solid (blue) lines respectively. \textbf{Right:} The posterior distribution and 95\% credible region for parameters $\Phi_o$ and $\gamma$ in Scenario VIb.}
	\label{fig:789-post}
\end{figure}

The cumulative mass profile credible regions from Group (3) ($\alpha$ free) are almost identical to those of Group (1) and so we do not bother showing them. Similar to the results in Section~\ref{sec:123}, the cumulative mass profiles between Scenarios VII, VIII, and IX are strikingly similar, regardless of the velocity anisotropy assumption. Overall, the estimate of $M_{125}$ is approximately $4.1\times10^{11}\msun$, with 50\% credible regions of about (3.6, 4.5) $\times10^{11}\msun$. The $M_{125}$ estimates are significantly lower than the estimates in Group (2), and only slightly lower than the estimates in Group (1).

The $\alpha$ estimates in Scenarios VII, VIII, and IX were all $\sim3.2$, with very narrow 50\% marginal credible intervals (Table~\ref{tab:summary}). This implies a slightly shallower tracer profile than we expected. It is interesting that the mass estimates are also lower in these Scenarios--- \cite{deason2012} noticed the same behaviour between $\alpha$ and the mass estimate. 

\begin{figure}[h]
	\centering
	\includegraphics[scale=0.45]{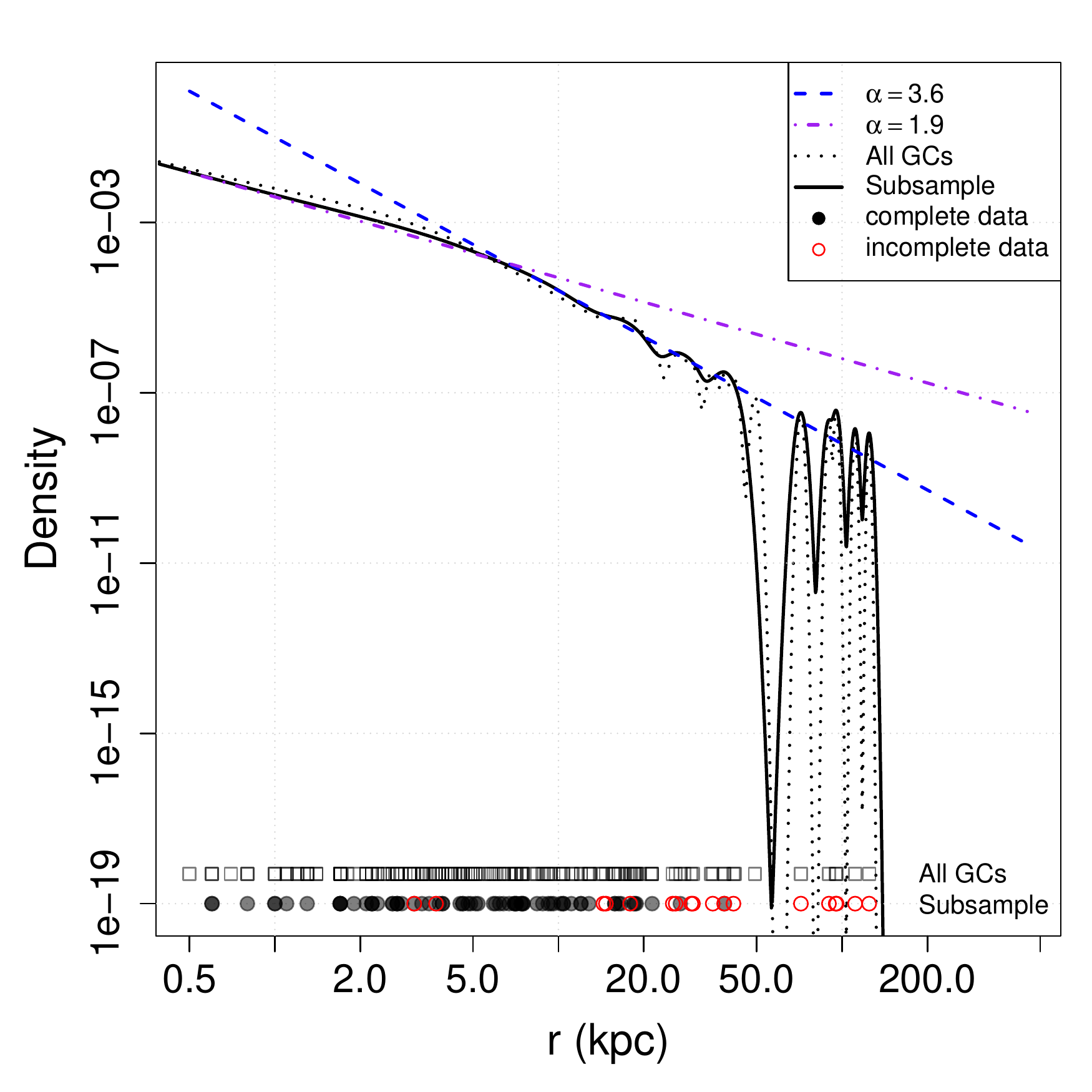}
	\caption{Smoothed radial density distribution of the GC subsample (solid line). Lines with power-law slopes 3.6 and 1.9 follow the outer and inner regions of GCs respectively. Points along the bottom of the graph indicate the $r$ values of individual GCs; the top row shows the entire GC population from Table~\ref{tab:deluxetable}, and the bottom row is the subsample used in our analysis. The lack of GCs around $r=50\kpc$~ results in the large drop in the density estimate in this region.}\label{fig:densGCs}
\end{figure}

One possible explanation for the lower $\alpha$ estimate is that the GCs in our subsample of Table~\ref{tab:deluxetable} do not follow a power-law profile with index $\alpha\approx3.5$. As a check, we calculate the spherical density profile for both the entire set of GCs from Table~\ref{tab:deluxetable} and the subset we used in our analysis (Figure~\ref{fig:densGCs}). The density profile of the subsample appears to follow a power-law slope of $3.6$ beyond $\sim7\kpc$~just as well as the entire data set. However, we can also see that a power-law slope of 1.9 follows the density profile quite well within $\sim6\kpc$. In our subsample, 42 of the 89 GCs fall within this inner region, so when $\alpha$ is free, the best model fit of a single power-law is a compromise of these two slopes. A broken power-law might be a better description of the GC profile in the Milky Way. Alternatively, we could exclude data within 6.64\kpc~(the point of intersection of the two straight lines shown in Figure~\ref{fig:densGCs}) at the expense of a smaller sample size of tracers.

To test the latter hypothesis, we run Scenario IX again but only use data for which $r>6.64\kpc$. We also re-define the prior distribution on $\alpha$ for consistency, using $b=6.64\kpc$ and only using the extra data beyond 6.64\kpc. While the resulting estimate of $M_{125}$ is larger ($5.06\times10^{11}\msun$, with 50\% credible region 4.23-5.71$\times10^{11}\msun$), the estimates for $\alpha$ and $\beta$ are relatively unchanged (3.14 and 0.33 respectively). Therefore, excluding inner region objects leads to a slightly higher mass estimate despite an unchanged $\alpha$ estimate. We return to this point in more detail in Section~\ref{sec:rcut}.

\subsection{Group (4): Scenarios X, XI, \& XII}\label{sec:10}

The mean $M_{125}$ estimates for Group (4) are lower than those in Group (2), but higher than those in Groups (1) and (3) (Table~\ref{tab:summary} and Figure~\ref{fig:Mresults}). However, Scenario X and XI's 95\% credible regions for the mass estimate overlap with the mass estimates from all Groups (Figure~\ref{fig:Mresults}).

\begin{figure}
\centering
	\includegraphics[scale=0.32]{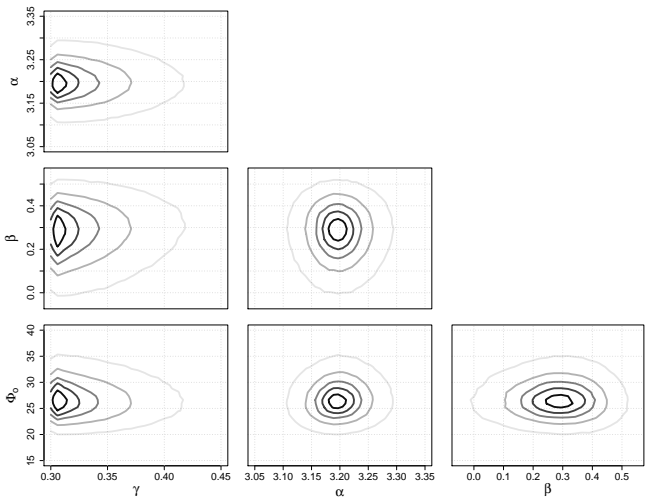}
	\caption{\textbf{Posterior distributions for Scenario XII.} The 10, 30, 50, 75, and 95\% joint credible regions are shown as contours.}
	\label{fig:Xposteriors}
\end{figure}

\begin{figure*}
\centering
	\includegraphics[ scale=0.5]{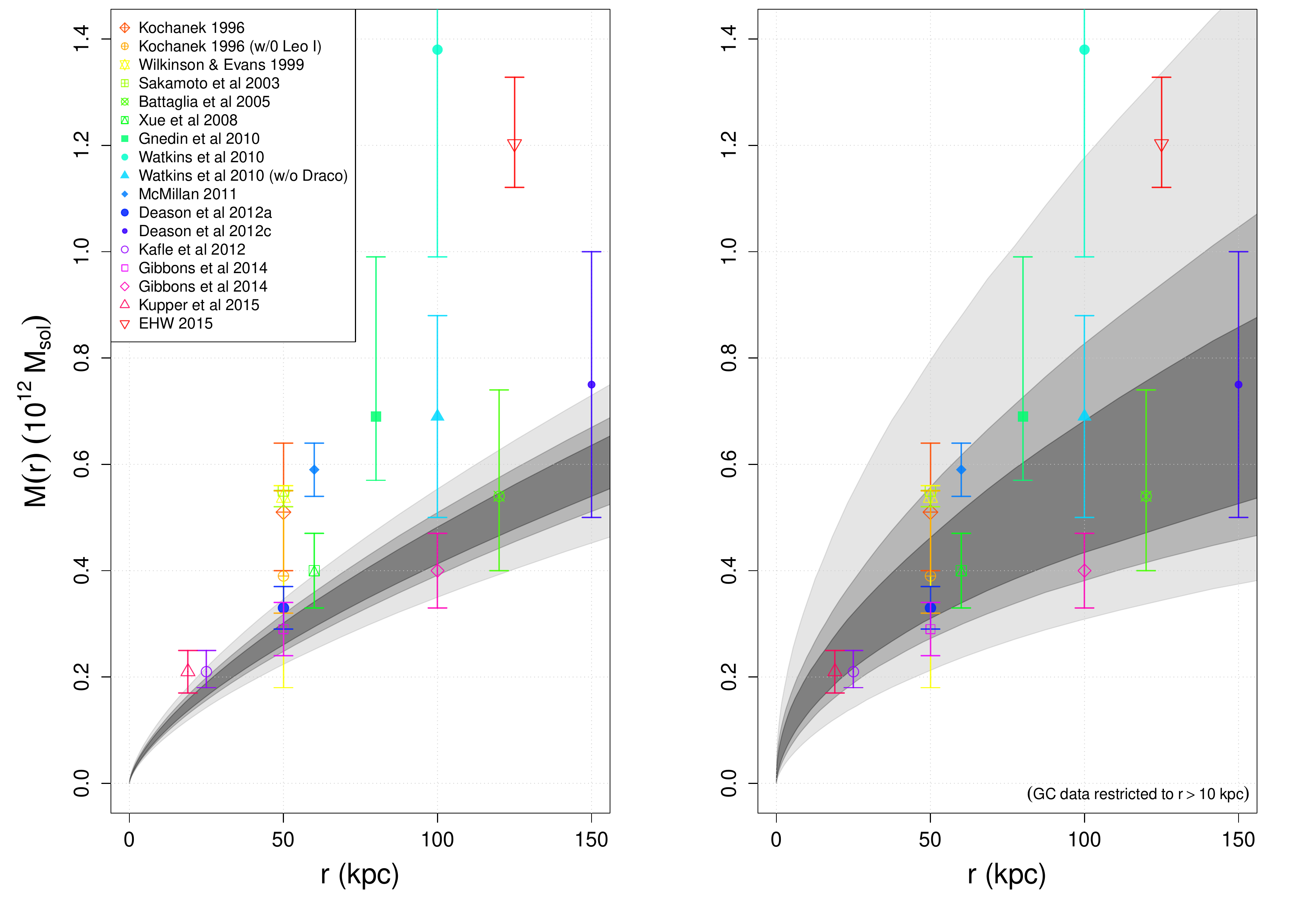}
	\caption{\textbf{Comparisons of Scenario XII $M(R<r)$ profile credible regions to mass estimates from other studies} \citep[those not mentioned in the text are][]{ kochanek1996, mcmillan2011, kafle2012, 2015EHW, kupper2015}. The shades of grey (dark to light) correspond to the Bayesian credible regions (95, 75, and 50\%). \emph{Left:} Mass profile when all GC data are used in the analysis. \emph{Right:} Mass profile when GC data at Galactocentric positions $r<10$\kpc~ are excluded.}
	\label{fig:Xresults-compare}
\end{figure*}

Scenario XII deserves some attention, as all four model parameters are free. The joint posterior credible regions for all parameter combinations are shown in Figure~\ref{fig:Xposteriors}, the parameter estimates are shown in Table~\ref{tab:summary}, and the cumulative mass profile is on the left-hand side of Figure~\ref{fig:Xresults-compare}.

One notable feature in Figure~\ref{fig:Xposteriors} is in the marginal posterior distributions for $\gamma$ (the leftmost column in the figure). The mode for $\gamma$ is very close to the lower bound defined in the prior distribution, similar to the situation seen in Group (2) (Figure~\ref{fig:456b-post}). The posterior distribution for $\gamma$ also has an asymmetric shape, reminiscent of the shape seen in Figure~\ref{fig:2dpost}. The parameter $\gamma$ was poorly constrained when we used a less informative prior (Figure~\ref{fig:2dpost}), and appears to be constrained in Group (2) and (4) mainly because of the more informative prior. Therefore, the present GC sample may not provide enough information about the dark matter halo to constrain its shape, without making relatively strong prior assumptions. The most we can say is that perhaps the potential is shallower than NFW.

The other notable feature about Scenario XII is that the 95\% credible regions for $M_{125}$ overlap the 50\% credible regions from all the other Scenarios.

The points in Figure~\ref{fig:Xresults-compare} are results from other studies, which are discussed below (Section~\ref{sec:discussion}). In general, however, it is clear that the results of Scenario XII are in agreement with studies that favour a ``lighter'' dark matter halo.

\subsection{Summary of Groups (1) - (4)}

Figure~\ref{fig:Mresults} shows the effect that parameter assumptions can have on mass estimates. In Group (1) --- Scenarios I, II, and III--- the only parameter that was allowed to vary in every case was $\Phi_o$. The mass estimates are in very good agreement with one another, despite the different assumptions of velocity anisotropy. 

In Group (2) ($\gamma$ free) the mass estimates are consistently higher than Groups (1) and (3), but the uncertainties in these estimates are also substantially larger. Likewise, the credible regions for the $M(r)$ profiles  are much larger (Figure~\ref{fig:Mprofiles456}). This may imply that it is difficult to constrain the steepness of the dark matter profile using the kinematic information of the tracers, most of which lie within 30\kpc.  When a strong assumption is made about the dark matter potential shape ($\gamma=0.5$), the mass profile is more constrained but may be biased.  When we relax the restriction on $\gamma$, the mass becomes more uncertain and relies more heavily on the prior distribution $p(\gamma)$, again implying that the current sample of GCs cannot constrain $\gamma$ well on their own.

In Group (3) ($\alpha$ free), the mass estimates are lower than in any other group, but the results of Scenarios VII, VIII, and IX are in agreement with each other. We also observe that the tracer number density parameter $\alpha$ has a minor effect on the $M_{125}$ estimate, despite its lack of appearance in Equation~\ref{eq:Mr}.

The mass estimates in Group (4) are slightly lower than those in Group (2), but higher than Groups (1) and (3). The uncertainties in Group (4) are similar to those seen in Group (2), where the $\beta-$free case has smaller credible regions. This is attributed to allowing $\beta$ to be a free parameter, which narrows the allowed values of $\Phi_o$ via the likelihood, as shown in Section~\ref{sec:456}. In Groups (3) and (4) ($\alpha$ free), the mass estimates are lower than in Groups (1) and (2) respectively. The lower estimates of $\alpha$ in comparison to a fixed $\alpha=3.5$ (Table~\ref{tab:summary}) may be causing the lower mass estimate, a relationship also noted by \cite{deason2012}.

Table~\ref{tab:Scenarios} shows the anisotropy parameter estimates for the Scenarios in which $\beta$ was a free parameter (Scenarios III, VIb, IX, and XII). All estimates of $\beta$ are in agreement with one another within the 95\% credible intervals despite the mass estimates for these Scenarios differing in a large way. For a direct comparison of the anisotropy estimates with observations, we also estimate $\beta$ directly from the \emph{complete} data (71 GCs have both radial and tangential velocity measurements), and \textbf{obtain a mean value $\langle \beta \rangle=0.209$}. This value is slightly but not dramatically smaller than the values in Table~\ref{tab:Scenarios}, and falls within our 95\% credible regions of the $\beta$ estimates. 

Taking all considerations into account, we choose Scenario XII as our estimate for the Milky Way dark matter halo. Using the posterior distribution from Scenario XII, and assuming a Hubble constant of 67.8\kms$\text{Mpc}^{-1}$ \citep{Planck2015}, we extrapolate out to the virial radius defined by $\rho_{200}=200\rho_{crit}$. We find $r_{vir}=185^{+7}_{-7}$~\kpc, and obtain a virial mass for the Milky Way of $6.82\times10^{11}$\msun with 50\% credible region of $(6.06, 7.53) \times 10^{11}$\msun.


\subsection{Sensitivity Test of the GC Sample}\label{sec:rcut}
In this section, we run a simple sensitivity test of the relative importance of the inner vs. outer GCs. Although the GCs cover a range of $0.6<r<125$\kpc, many of the GCs in our sample are within $r=30\kpc$. Setting $\gamma=0.5$ is akin to assuming an NFW potential beyond $\simeq 10\kpc$. However, a single power-law will not account for the inner DM halo which is presumably flatter than the outskirts. Furthermore, the inner GCs ($r<10kpc$) are in a region where the bulge and disk are important contributors to the gravitational potential.

\begin{figure}[h]
	\centering
	\includegraphics[trim=0cm 0cm 0cm 1.5cm, scale=0.49]{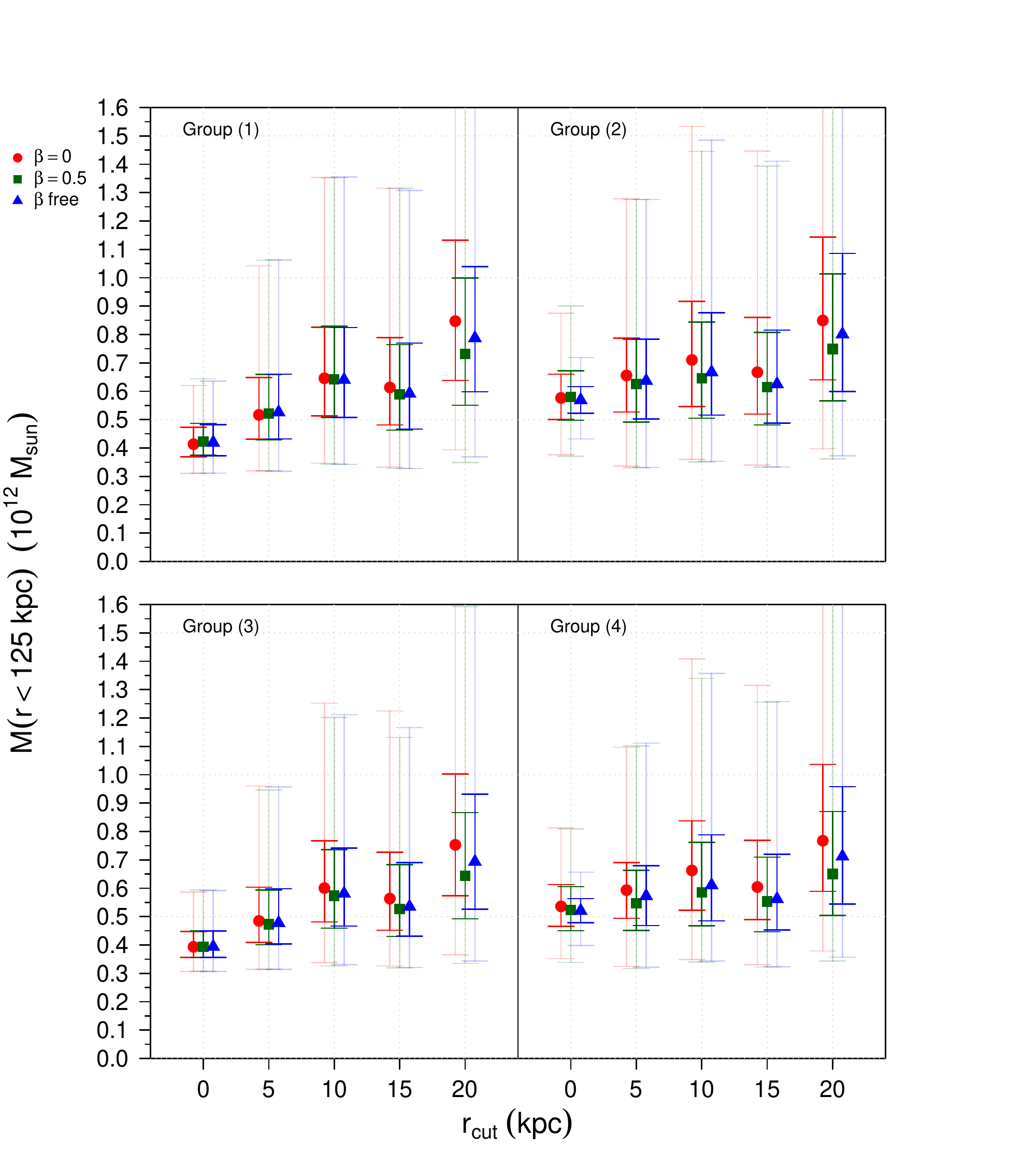}
	\caption{\textbf{Results from sensitivity tests; dependency of the $M_{125}$ estimates on the GC data sample.} Each mass estimate was determined using GCs beyond $r_{\text{cut}}$. Bright and faint error bars correspond to the 50\% and 95\% credible regions respectively.}
	\label{fig:rcutGrid}
\end{figure}

To investigate possible biases in our results, we perform a simple empirical test that has little reliance on a particular form (such as NFW) for the shape of the potential. We repeat the analysis for the whole suite of investigations (Table~\ref{tab:Scenarios}) using four different $r$ cut-off values, $r_{\text{cut}}$, for the data. Mass estimates within 125\kpc~ are obtained when GCs within 5,10,15, and 20\kpc~are excluded from the analysis respectively. We expect that as $r_{\text{cut}}$ becomes larger, the uncertainties in the mass estimate will increase simply due to the lower number of data points. The question remains whether or not there is a significant trend towards a lower or higher mass as $r_{\text{cut}}$ increases.

The number of data points decreases quickly as $r_{\text{cut}}$ increases. For $r_{\text{cut}}=5$\kpc, only 53 GCs remain in the sample (down from 89). For $r_{\text{cut}}$ values of 10, 15 and 20\kpc, the number of GCs drops to 33, 24, and 16 respectively. Furthermore, as $r_{\text{cut}}$ increases from 5 to 20\kpc, the percentage of \emph{incomplete} data increases from 30.2\% to 81.2\%, making the estimates of $\beta$ in Scenarios III, VI, IX, and XII more uncertain.

Figure~\ref{fig:rcutGrid} shows the median mass estimates within 125\kpc~ for all Scenarios listed in Table~\ref{tab:Scenarios}. Note that in all Groups, the Bayesian credible regions increase significantly as more data is excluded, as we expected. In Groups (1) and (3), when $\gamma=0.5$, there is a noticeable dependence on the inner GC data that causes lower mass estimates; the mass estimates trend to higher values as more inner GCs are excluded from the analysis. This dependence, however, is not as strong when the 95\% Bayesian credible regions are taken into account (the faint error bars in Figure~\ref{fig:rcutGrid}).

In contrast, for Groups (2) and (4), where $\gamma$ is a free parameter, the trend of increasing mass estimates as $r_{\text{cut}}$ increases is lessened significantly. Closer examination of Figure~\ref{fig:rcutGrid} reveals that the biggest mass differences between Groups occurs when all of the data is used in the analysis. On the other hand, there is little variation between the Groups' mass estimates when $r_{\text{cut}}>10$\kpc. Thus, the best choice may be to treat $\gamma$ as a free parameter, and exclude GCs within 10\kpc.

The estimates of $\beta$ when it is a free parameter (i.e. Scenarios III, VI, IX, and XII) do not depend heavily on the $r_{\text{cut}}$ values, despite the percentage of incomplete data increasing as $r_{\text{cut}}$ increases. For example, the 50\% credible regions for $\beta$ in Scenario III, in increasing order of $r_{\text{cut}}$, are (0.28, 0.44), (0.21, 0.44), (0.34, 0.58), and (0.01, 0.49).

Another interesting feature of Figure~\ref{fig:rcutGrid} is that as $r_{\text{cut}}$ increases, the difference in mass estimates between anisotropy scenarios ($\beta=0,0.5$ or free) becomes more pronounced. This can be attributed to the lowered percentage of complete data in the analysis as inner GCs are excluded. As the percentage of incomplete data becomes higher, any assumptions about $\beta$ will have a larger impact on the mass estimate.

\section{Discussion \& Future Work}\label{sec:discussion}

The mass estimate for the Galaxy in this work is significantly lower than the estimate from EHW's analysis (Figure~\ref{fig:Xresults-compare}). However, EHW  used two populations of satellites (DGs and GCs) and assumed a self-consistent Hernquist model. When we apply the isotropic Hernquist model to the data used in this paper (GCs only), the results are in closest agreement with Group (3); \textbf{the isotropic Hernquist model gives a mass interior to 125\kpc~of $3.74 (3.64, 3.81) \times 10^{11}\msun$}. Thus, the biggest difference results from dropping the DGs, which populate $r \gtrsim 50$\kpc.

In general, our results are closely consistent with a number of other studies which suggest a ``light'' Milky Way total mass. We take Scenario XII as our best estimate for the Milky Way mass profile (Figure~\ref{fig:Xresults-compare}) and compare it to other studies. We consider the results when all GC data are used, and when only GCs at $r>10\kpc$ are included. It is clear from Figure~\ref{fig:Xresults-compare} that the latter case results in better agreement with more studies, simply because of the increased range of the Bayesian credible regions at each radius.

We now discuss and compare our results from Scenario XII with a selection of the mass estimates shown in Figure~\ref{fig:Xresults-compare} (left). We also discuss some other studies whose results cannot be displayed in the figure.

\cite{wilkinsonevans1999} used kinematic data of GCs and DGs, and a truncated flat-rotation curve (TF) model, to estimate the mass of the Milky Way. Their result for $M(r<50\kpc)$ and the uncertainties are shown in Figure~\ref{fig:Xresults-compare}. The lower bound of their uncertainty completely overlaps our mass estimate at $50\kpc$, despite the difference in model assumption. The point estimate is not in good agreement with ours, but this may be attributed to the different model choice. 

Our results are in good agreement with \cite{sakamoto2003}, who used satellite galaxies, GCs, and field horizontal branch stars as kinematic tracers and found $(M(r<50\kpc)=5.5^{+0.0}_{-0.2}\times10^{11}\msun$.

\cite{battaglia2005} studied the radial velocity dispersion profile of 240 halo objects (satellite galaxies, GCs, and halo stars), and found that an NFW model predicts a virial mass of $0.8^{+1.2}_{-0.2}\times10^{12}$\msun with a concentration of $c=18$. The virial mass is an extrapolation beyond their furthest data point ($120\kpc$), so they also quote their best mass estimate $M(r<120kpc)=5.4^{+2.0}_{-1.4}\times10^{11}$\msun. Figure~\ref{fig:Xresults-compare} shows that their result is in excellent agreement with our estimate at 120\kpc.

\cite{xue2008} compared the line-of-sight velocity distribution of 2401 BHB stars to cosmological galaxy simulations to constrain the mass of the Milky Way. They found an enclosed mass within 60\kpc~of $4.0\pm0.7\times10^{11}\msun$, in agreement with our Scenario XII results (see Figure~\ref{fig:Xresults-compare}). However, when they assume an NFW halo profile the virial mass estimate is $1.0^{+0.3}_{-0.2}\times10^{12}\msun$, which is not in agreement with $M_{vir}$ presented here (see Section~\ref{sec:10}). This disagreement cannot be attributed to the dark matter profile parameter $\gamma$, since we found that a lower $\gamma$ estimate leads to a higher mass estimate (Section~\ref{sec:456}), and we found $\gamma \approx 0.3$, rather than the NFW approximation of $\gamma=0.5$, 

\cite{li2008} report a virial mass of $2.43\times10^{12}$\msun, with a lower limit of $0.8\times10^{12}$\msun at the 95\% confidence level. This is at the higher end of the results in the literature and disagrees with our results.

As mentioned in Section~\ref{sec:intro}, W10 calculated the Milky Way's mass with two different mass estimators, assuming an NFW-type halo and using kinematic data from 26 DGs. Their mass estimates depended significantly on both the velocity anisotropy assumption for the tracers and the inclusion (or not) of proper motion measurements; they reported $M(r<300\kpc)$ ($M_{300}$) values ranging from 7.0 to $14.0\times10^{11}\msun$. To compare, we extrapolate the Group (3) analyses (i.e. those which assume an NFW-type halo) out to $M_{300}$ and find a range of 4.8 to 9.2$\times10^{11}\msun$ for the 95\% credible intervals, independent of the velocity anisotropy. Although our estimates were made using GCs rather than DGs, the 95\% credible regions do overlap with the lower end of the mass estimates from W10. If instead we compare our Scenario XII results extrapolated out to $M_{300}$, we obtain a 95\% credible range of (6.84,11.93)$\times10^{11}\msun$, which is in much better agreement with W10's estimates. However, it should be noted that in Scenario XII, the posterior distribution for $\gamma$ does not suggest an NFW-type potential, which \emph{was} assumed by W10.

Using distant halo stars and a high-velocity star sample, and the spherical Jeans equation, \cite{gnedin2010} found an enclosed mass for the Milky Way of $M(80\kpc)=6.9^{+3.0}_{-1.2}\times10^{11}\msun$. This is in disagreement with Scenario XII's estimate of $M(80\kpc)$ (see Figure~\ref{fig:Xresults-compare}), but in better agreement with Scenarios IVb, Vb, and VIb (refer to Figure~\ref{fig:456b-post}).

\cite{busha2011} also used Bayesian inference to estimate the mass of the Milky Way, but instead incorporated \LCDM -based simulations and the phase-space information of the Small and Large Magellanic Clouds. They arrived at a virial mass estimate of $ 1.2^{+0.7}_{-0.4}\times10^{12}\msun$, where we quote only the statistical errors. Their virial radius was $250^{+60}_{-30}\kpc$. Our Scenario XII estimates within 250\kpc~and 310\kpc~are in close agreement: $0.83~(0.61, 1.05) \times10^{12}$\msun and $0.96~(0.70, 1.22)\times10^{12}$\msun.

\cite{deason2012} used the power-law model employed here in a maximum likelihood analysis of BHB star kinematics. When they assume spherical symmetry and set $\alpha=3.5$, their $\gamma$, $\Phi_o$, and $\beta$ values were $0.35_{-0.17}^{+0.08}$,
 $30\pm5 \times 10^4\text{km}^2\text{s}^{-2}$, and $0.4_{+0.1}^{-0.2}$ respectively, which leads to mass estimate within 50\kpc~ of $3.3\pm0.4 \times 10^{11} \msun$. In our equivalent set-up (Scenario VIb), we estimated  $\gamma \approx 0.33$, $\Phi_o \approx 30 $, $\beta \approx 0.27$, and a mass within 50\kpc~ of $3.09 ~(2.43, 3.86) \times 10^{11} \msun$, which is in very good agreement despite the use of different tracer objects and a different method.

Using the TME from W10, \cite{deasonetal2012MNRAS} concluded that the mass within 150kpc is (5-10)$\times10^{11}\msun$. When we extrapolate the Scenario VIb mass profile out to 150\kpc, we find the mass and 95\% credible regions to be $6.44\times10^{11}\msun$, (4.84,8.14)$\times10^{11}\msun$, again in good agreement. In Scenario XII the estimate with 95\% credible regions is $5.90~(4.45, 7.43) \times10^{11}$\msun, which is also in good agreement with the TME. Again, this suggests a ``light'' Milky Way total mass.

The virial mass of the Milky Way was estimated by \cite{boylan2013} to be $1.6^{+1.0}_{-0.6} \times 10^{12}$\msun, where the uncertainties represent 90\% confidence intervals and the virial radius was $\sim 300\kpc$. The 95\% credible regions for $M(r<300\kpc)$ from Scenario XII are in agreement at (0.68,1.19)$\times10^{12}$\msun.

Lastly, \cite{gibbons2014} used a Bayesian method and kinematic data from the Sagittarius stream to measure the Galactic mass distribution, and reported the mass within 50 and 100 kpc: $M(50\kpc)=2.9\pm0.5 \times 10^{11} \msun$ and $M(100\kpc)=4.0\pm0.7\times10^{11}\msun$. The former is in agreement with our Scenario XII estimate for $M(50\kpc)$ mentioned above, and is also in reasonable agreement with $M(100\kpc)$ (see Figure~\ref{fig:Xresults-compare}). Their result is also in line with the aforementioned papers that support a lighter Milky Way Galaxy. \cite{gibbons2014} also point out that their leaner mass estimate of the Milky Way helps to solve the ``Too big to fail'' problem.

In Scenario XII, where the GC subsample is limited to $r>10\kpc$, the mass profile credible regions widen dramatically (right-hand side of Figure~\ref{fig:Xresults-compare}). Under this circumstance, our results agree with almost every value we have quoted from the literature.

There are some issues with the analysis presented here:
\begin{itemize}

\item{ We assumed a spherically symmetric DM halo. However, the geometry of the Milky Way's DM halo is not well known. Some studies favour a triaxial shape or oblate shape \citep{loebman2014ApJ, deg2013MNRAS,lawmajewski2010ApJ}, others a prolate shape \citep{bowden2016MNRAS}, and still others show that a spherical shape is not ruled out \citep{smith2009ApJ}. The ideal way to allow for a non-spherical halo under the methodology of EHW would be to use a DF that includes an angular-dependent dark matter potential through extra model parameters. These parameters could then be estimated via the posterior distribution, and used to make inference about the geometry of the dark matter halo.}

\item{The model we used here assumes a constant anisotropy of the tracers, not an anisotropy that can vary with distance.}

\item{The assumption that tracers are randomly distributed about the Galaxy is probably incorrect at large $r$. Substructure in the distribution of Galactic satellites and halo stars arises in many hierarchical formation simulations of Milky Way-type galaxies, is becoming increasingly obvious in M31 \cite[e.g.][]{mcconnachie2009nature,ibata2007ApJ}, and is without a doubt present in our own Galaxy \cite[e.g.][]{yanny2000, belokurov2006ApJ}. However, recently \cite{yencho2006} showed that the errors introduced by assuming a randomly distributed tracer population are actually quite small (at the 20\% level) in comparison to the errors introduced by measurement uncertainties. This brings us to the next point.}

\item{The mass estimate under the power-law model presented here depends on the range of GCs used in the analysis. To fully understand this dependence, a better test of this simple power-law form ($\gamma=$ constant) will be to apply this analysis to simulated galaxies built from \LCDM~ hydrodynamic simulations. This will be the subject of an upcoming paper (Eadie et al 2016, in preparation).
}

\item{Measurement uncertainties have not been included in this analysis, but as EHW showed they are extremely important. In the third paper of this series, we will discuss how uncertainties can be included in the Bayesian paradigm; the interested reader may look at some preliminary tests of this method in \cite{eadieJSM, eadieIAU}.}

\item{Finally, although our method does use incomplete and complete data at the same time, it relies on geometric arguments to incorporate the incomplete data (that is, the requirement that $|cos\xi|>0.95$). There are 157 GCs in Table~\ref{tab:deluxetable}, and we used only 89 of them ($\sim56$\%). In our next paper, we will show how this problem is remedied through the use of a hierarchical model that includes the measurements uncertainties.}

\end{itemize} 

The Bayesian analysis performed here highlights the important influence of parameter assumptions and selecting prior probabilities. The results of Scenarios IV, V, and VI, as well as IVb, Vb, and VIb in Section~\ref{sec:456} showed that the data cannot constrain the dark matter halo profile parameter $\gamma$ very well, without prior information from other studies and knowledge gained from simulations. However, this highlights a strength of the Bayesian approach--- we must mathematically and explicitly state our prior knowledge and assumptions. Many of these assumptions are incorporated into studies that use maximum likelihood and standard frequentist techniques too, but the assumptions may be more implicit.

\section{Conclusion}

We have performed a Bayesian analysis to determine the mass and cumulative mass profile of the MW out to 125kpc using GCs as tracers of the Galactic halo. The model and method we used is sensitive to the assumptions about model parameters, and most notably assumptions about the power-law profile of the dark matter halo potential. There also appears to be a dependence on the positions of the GCs, as excluding inner GCs leads to slightly higher mass estimates, especially when strong assumptions have been made about the dark matter profile.

One advantage of the EHW method is that we can easily obtain an estimate for the mass enclosed at any radius, and immediately obtain uncertainties in that estimate. This feature of the method makes comparing our results to other studies relatively straightforward (e.g. Figure~\ref{fig:Xresults-compare}). Another advantage is that using both complete and incomplete data simultaneously seems to remove any mass-anisotropy degeneracy. Furthermore, independent of parameter assumptions, the results suggest that the GC population has a mild radially anisotropic velocity distribution.

The first data release from the Gaia mission, including astrometric and photometric measurements, will occur in summer 2016 \citep{GAIAweb}. Tycho-Gaia astrometric solution (TGAS), which uses data from the Hipparcos catalog and the new measurements from GAIA, could yield proper motion, parallax, and position measurements for 2.5 million Tycho-2 stars \citep{michalik2015}. The analytical approach described here will be well suited to this data. 

Lastly, our analysis and comparison to other studies, and in particular Figure~\ref{fig:Xresults-compare}, strongly emphasizes the need for remote, virialized tracers ($r\gtrsim 30$kpc) with proper motion measurements to place stronger constraints on the slope of the dark-matter halo profile, and ultimately the total mass of the Milky Way. As more complete data at large Galactic distances become available, it will also be easier to rule out possible dark matter halo models via Bayesian model comparison tests.

\section*{Acknowledgements}
GME and WEH thank the anonymous referee for thoughtful criticisms and constructive feedback on this paper. GME also thanks Aaron Springford for extremely helpful input about conjugate prior distributions and Gandhali Joshi for some  helpful mathematical conversations. The authors also want to thank Alis Deason for the detailed and timely answers to our inquiries, and Dana Casetti for sharing the updated list of GC proper motions.

WEH and GME acknowledge the financial support of NSERC. 

\bibliographystyle{apj}

\bibliography{myrefs}

\label{lastpage}

\end{document}